\documentclass[modern]{aastex631}

\usepackage{graphicx}

\newcommand{\OneExp}{\textsf{OneExp}}
\newcommand{\TwoExp}{\textsf{TwoExp}}

\shorttitle{SN2SNR OneExp/TwoExp}
\shortauthors{Ferrand et al.}

\begin{document}

\title{The role of the secondary white dwarf in a double-degenerate double-detonation explosion, in the supernova remnant phase}

\author[0000-0002-4231-8717]{Gilles Ferrand}
\newcommand{\UM}{The University of Manitoba, Department of Physics and Astronomy, Winnipeg, Manitoba, R3T~2N2, Canada}
\newcommand{\iTHEMS}{RIKEN Center for Interdisciplinary Theoretical and Mathematical Sciences (iTHEMS), Wak\={o}, Saitama, 351-0198 Japan}
\affiliation{\UM}
\affiliation{\iTHEMS}

\author[0000-0003-3308-2420]{R\"udiger Pakmor}
\affiliation{Max-Planck-Institut f\"{u}r Astrophysik, Karl-Schwarzschild-Str. 1, D-85748, Garching, Germany}

\author{Yusei Fujimaru}
\newcommand{\KyotoU}{Department of Astronomy, Kyoto University, Kitashirakawa, Oiwake-cho, Sakyo-ku, Kyoto 606-8502, Japan}
\affiliation{\KyotoU}

\author[0000-0002-2899-4241]{Shiu-Hang Lee}
\affiliation{\KyotoU}
\affiliation{Kavli Institute for the Physics and Mathematics of the Universe (WPI), The University of Tokyo, Kashiwa 277-8583, Japan}

\author[0000-0001-6189-7665]{Samar Safi-Harb}
\affiliation{\UM}

\author[0000-0002-7025-284X]{Shigehiro Nagataki}
\newcommand{\ABBL}{Astrophysical Big Bang Laboratory (ABBL), RIKEN Pioneering Research Institute (PRI), Wak\={o}, Saitama, 351-0198 Japan}
\newcommand{\RBCiTHEMS}{RIKEN-Berkeley Center, RIKEN iTHEMS, University of California, Berkeley, Berkeley, CA 94720, USA}
\newcommand{\ABBG}{Astrophysical Big Bang Group (ABBG), Okinawa Institute of Science and Technology Graduate University (OIST), 1919-1 Tancha, Onna-son, Kunigami-gun, Okinawa 904-0495, Japan}
\affiliation{\ABBL}
\affiliation{\iTHEMS}
\affiliation{\RBCiTHEMS}
\affiliation{\ABBG}

\author[0000-0002-4460-0097]{Friedrich K.\ R{\"o}pke}
\affiliation{Heidelberger Institut f{\"u}r Theoretische Studien, Schloss-Wolfsbrunnenweg 35, 69118 Heidelberg, Germany}
\affiliation{Zentrum f{\"u}r Astronomie der Universit{\"a}t Heidelberg, Institut f{\"u}r Theoretische Astrophysik, Philosophenweg 12, 60120 Heidelberg, Germany}
\affiliation{Zentrum f{\"u}r Astronomie der Universit{\"a}t Heidelberg, Astronomisches Rechen-Institut, M{\"o}nchhofstr.\ 12--14, 69120 Heidelberg, Germany}

\author[0000-0002-1796-758X]{Anne Decourchelle}
\affiliation{Université Paris-Saclay, Université Paris Cité, CEA, CNRS, AIM, 91191, Gif-sur-Yvette, France}

\author[0000-0002-5044-2988]{Ivo R. Seitenzahl}
\affiliation{Research School of Astronomy and Astrophysics, Australian National University, Canberra, ACT 2611, Australia}
\affiliation{Mathematical Sciences Institute, Australian National University, Canberra, ACT 0200, Australia}

\author[0000-0002-7507-8115]{Daniel Patnaude}
\affil{Smithsonian Astrophysical Observatory, 60 Garden Street, Cambridge, MA 02138 USA}

\begin{abstract}

Type Ia supernovae (SNe) are believed to be thermonuclear explosions of white dwarf (WD) stars, but their progenitor systems and explosion mechanisms are still unclear. Here we focus on double degenerate systems, where two WDs are interacting, and on the double detonation mechanism, where a~detonation of a helium shell triggers a detonation of the carbon-oxygen core of the primary WD. We take the results from three-dimensional SN simulations of \cite{Pakmor2022FateSecondary} and carry them into the supernova remnant (SNR) phase, until 1500~yr after the explosion. We reveal signatures of the SN imprinted in the SNR morphology. We confirm the impact of a companion on the SNR: its presence induces a conical shadow in the ejecta, that is long lived. Its intersection with the shocked shell is visible in projection as a ring, an ellipse, or a bar, depending on the orientation. New, we test the case of a nested explosion model, in which the explosion of the primary induces the secondary to also explode. As the explosion of the secondary WD is weaker only the primary outer ejecta interact with the ambient medium and form the main SNR shell. The secondary inner ejecta collide with the reverse shock, which enhances the density and thus the X-ray emissivity. The composition at the points of impact is peculiar, since what is revealed are the outer layers from the inner ejecta. This effect can be probed with spatially-resolved X-ray spectroscopy of young SNRs. 
\end{abstract}


\section{Introduction} 
\label{sec:intro}

Type Ia supernovae (SNe) are understood as the runaway thermonuclear explosion of a degenerate white dwarf (WD) in a close binary system. At cosmological scales, they provide us with a way of measuring distances, and were used to establish the acceleration of the expansion of the universe \citep{Riess1998AcceleratingUniverse,Perlmutter1999MeasurementsSupernovae}. At galactic scales, they are major sites for nucleosynthesis and contribute to the enrichment of the interstellar medium (ISM) with heavy elements \citep{Seitenzahl2017NucleosynthesisThermonuclearSupernovae}. Despite their prominence in astrophysics there remain many open questions about their origin, see \cite{LiuRoepke2023TyepIaReview} and \cite{RuiterSeitenzahl2025TypeIapuzzle} for recent reviews. One long standing puzzle is the nature of the progenitor system(s), which may comprise one WD and a normal star or two WDs. Another issue, that has lead to an abundance of theoretical models, is the explosion mechanism for the WD, which depends on whether the WD's mass is near or lower than the Chandrasekhar mass. 
In this work we focus on the double degenerate scenario where two WDs are interacting, and on the double detonation mechanism involving a~WD explosion in two steps. 

The double degenerate scenario (\citealt{Iben1984Supernovaemass,Webbink1984DoubleSupernovae}) is consistent with the observed rate of normal Type Ia SNe (\citealt{Ruiter2009RatesSupernovae}), and it can reproduce the brightness distribution of normal SN Ia -- under the assumption that the brightness of the explosion is set by only the primary (\citealt{Ruiter2013Onthebrightnessmergers,Sato2016TheCriticalMassRatio}). It also explains the common lack of a companion or signatures from a companion, in SNe and their remnants (\citealt{RuizLapuente2019Survivingcompanions,LiuRoepke2023TyepIaReview}). 

The double detonation mechanism (\citealt{Fink2010Double-detonationCore,Livne1990SuccessiveSupernovae,Guillochon2010SurfaceDetonations,Pakmor2013Helium-ignitedSupernovae}) is a promising model for a binary WD. In this model the explosion is triggered by unstable dynamical accretion from the secondary WD onto the primary WD, on their way to but before the merger -- this is the ``dynamically-driven'' part of the D$^6$ moniker from \cite{Shen2018ThreeSupernovae}. Specifically, this explosion mechanism works in two steps: the accretion stream from the secondary induces a helium detonation in the outer shell of the primary, which engulfs the WD and drives a shock wave that converges in the core, where it ignites a carbon detonation that explodes the WD. The explosion in the double detonation scenario occurs many orbits before the system would merge. So in contrast to the violent merger scenario where the secondary WD is on its last orbit and already being disrupted at the time of explosion \citep{Pakmor2012NormalBinaries}, here the secondary WD is still intact \citep{Pakmor2022FateSecondary}. The recent detection of a double shell of calcium in a young supernova remnant (SNR) was taken as evidence that double-detonations of WDs indeed occur in nature \citep{Das2025Calcium}.

When the primary WD explodes, the secondary WD is still intact, and it has generally been assumed to survive (this is the case for instance in the simulations by \citealt{Pakmor2013Helium-ignitedSupernovae} and \citealt{Tanikawa2018Three-DimensionalCompanion}). But the fate of the secondary WD is uncertain. \cite{Tanikawa2019Double-DetonationMaterials} found that only the primary WD explodes in some double degenerate systems, while in others the secondary WD explodes as well, via a single detonation or via a double detonation like the above -- hence a range of double, triple, and quadruple explosion scenarios. This variety was also observed in the 2D simulations of \cite{Boos2024TypeIaBinary}. \cite{Pakmor2022FateSecondary} compared, for a given double degenerate system, the case when the secondary does not explode (one WD explosion, hereafter the \OneExp\ model) and the case when the secondary does explode (two WD explosions, hereafter the \TwoExp\ model). They focused on the early SN phase, and analyzed and interpreted the differences in the ejecta and their synthetic observables.

The aim of this work is to determine the different observational imprints produced by these two scenarios in the SNR phase, up to 1500 years post the explosion. The SNR is made by the interaction between the ejecta and the circumstellar and interstellar matter. During this phase, the ejecta are shocked (from the outside in) to X-ray emitting temperatures. The strong shocks bounding the interaction region can accelerate particles, that shine in radio and in $\gamma$-rays, and commonly in synchrotron X-rays filaments along the shock. The thermal X-ray emission offers critical diagnostics of the elemental composition and thermodynamical state of the (shocked) ejecta, and allows for a detailed mapping of their morphology and expansion (\citealt{Vink2012SupernovaPerspective,Reynolds2017DynamicalRemnants}). We can thus use the SNR as a probe of SN physics, provided we observe it young enough, before too much interaction with a complex ISM. 

This paper is the fourth in a series where we are investigating, by the means of hydrodynamic simulations, the morphological signatures imprinted on the SNR by the progenitor system and the explosion mechanism. In \cite{Ferrand2019FromExplosion} we showed how and until what age the imprint of the explosion can be detected, before hydrodynamic instabilities from the SNR phase dominate, for the case of a single Chandrasekhar-mass WD. In \cite{Ferrand2021FromModels} we showed how variants of the same explosion model, in terms of geometry of the ignition and propagation of the flame, can be distinguished in the morphology of the young SNR. In \cite{Ferrand2022_D6} we looked at a different kind of model, the D$^6$ model from \cite{Tanikawa2018Three-DimensionalCompanion} where the primary WD explodes and the secondary WD survives, and we observed the marked impact of the companion. Here, following up on \cite{Pakmor2022FateSecondary} we study the fate of the secondary WD, and for the first time we present the impact on a young SNR of a double stellar explosion: an explosion within an explosion.

The paper is organized as follows. In Section~\ref{sec:method} we summarize the method for carrying out the numerical simulations from the SN phase into the SNR phase. In Section~\ref{sec:results} we present the results on the SNR evolution, regarding the overall morphology and the distribution of elements as can be seen in X-rays. In Section~\ref{sec:discussion} we discuss implications of our findings for double detonation models and their impact on the interpretation of observations. In Section~\ref{sec:conclusion} we conclude and outline our perspectives.

\section{Method} 
\label{sec:method}

Our simulations are done in two steps, first the SN explosion, published in \cite{Pakmor2022FateSecondary}, then the SNR evolution, presented in this paper.

\subsection{SN model}

 The SN simulation pipeline of \cite{Pakmor2022FateSecondary} consists of four parts. (i)~Generate 1D profiles of two WDs in hydrostatic equilibrium. (ii)~Map these profiles onto the grid of a three-dimensional (3D) code and simulate the inspiral and mass transfer through the explosion and until the ejecta are in homologous expansion. This is done using the moving-mesh code \textsc{Arepo} \citep{Weinberger2020AREPOPublicCodeRelease}, that implements a quasi-Lagrangian finite volume method. The simulation takes into account both hydro and gravity, uses the Helmholtz equation of state \citep{TimmesSwesty2000HelmholtzEOS}, and is coupled with a 55-isotope nuclear reaction network. (iii)~Post-process the hydro run with tracer particles to obtain the detailed isotopic abundances; this time a 384-isotope nuclear reaction network is used. (iv)~Run a Monte-Carlo radiative transfer code to compute synthetic light curves and spectra \citep{Pollin2024syntheticobservables}. In this paper we are taking the simulation at the end of stage~3, when we have freely expanding ejecta with known elemental abundances on a 3D grid. 

\subsubsection{Basic properties of the explosion}
We summarize the basic properties of the SN model. The initial masses of the WDs are $M_1 = 1.05\,M_\odot$ and $M_2 = 0.7\,M_\odot=0.67\,M_1$, their central densities are $\rho_1=4.8\times10^7\,\mathrm{g}\,\mathrm{cm}^{-3}$ and $\rho_2=6.3\times10^6\,\mathrm{g}\,\mathrm{cm}^{-3}$, where everywhere subscripts 1 and 2 refer to respectively the primary and the secondary. Both WDs have the same composition, 50\% C and 50\% O in the core, plus an outer layer of $0.03\,M_\odot$ of pure He (this is a simplification, and a rather optimistic configuration that makes it easier to get the WDs to explode). 

In the 3D hydro simulation, the helium shell detonation of the primary happens at $t_{1,{\rm He}}=148.8\,s$, at the point where the accretion stream hits its surface. It wraps around the WD and sends a shockwave into its core, and the carbon detonation happens at $t_{1,{\rm C}}=t_{1,{\rm He}}+1.3$~s almost at the center (25\,km away from it, less than a percent of its radius), this second detonation burns the whole primary WD. The shockwave from the primary explosion hits the secondary WD, and ignites its helium shell at $t_{2,{\rm He}}=t_{1,{\rm C}}+0.9\,s=t_{1,{\rm He}}+2.2\,s$. The detonation converges at the core at $t_{2,{\rm C}}=t_{2,{\rm He}}+2.7\,s=t_{1,{\rm C}}+3.6\,s=t_{1,{\rm He}}+4.9\,s$, at that time the secondary is already completely engulfed by the expanding ejecta of the primary. In the base run this fails to ignite a carbon detonation, so only the primary is destroyed, this is the \OneExp\ case. The conditions are however probably suitable for a detonation, but not numerically resolved. Therefore, in a re-run, the second detonation is manually triggered by locally increasing the temperature. This leads to the complete destruction of the secondary and corresponds to the \TwoExp\ case. 

In terms of energy release, for the primary $E_{1,{\rm He}}=8.3\times10^{49}$~erg and $E_{1,{\rm C}}=1.4\times10^{51}$~erg, for the secondary $E_{2,{\rm He}}=6.7\times10^{49}$~erg and $E_{2,{\rm C}}=5.6\times10^{50}$~erg, so the carbon detonation(s) completely dominate(s) the energetics of the SN. The energy of the expanding ejecta is $E_{\rm sn,OneExp} = 1.4\times10^{51}$~erg versus $E_{\rm sn,TwoExp} = 1.9\times10^{51}$~erg $=1.36\times E_{\rm sn,OneExp}$. The mass of the ejecta is $M_{\rm ej,OneExp}=1.09\,M_\odot=0.79\,M_{\rm ch}$ (the primary and some part of the secondary) versus $M_{\rm ej,TwoExp}=1.75\,M_\odot=1.27\,M_{\rm ch} = 1.61\times M_{\rm ej,OneExp}$ (the full mass of both WDs). 

The \OneExp\ case, with one double detonation, is similar to the D$^6$ model from \cite{Tanikawa2018Three-DimensionalCompanion}. For reference, the latter has $M_1 = 1.0\,M_\odot$ and $M_2 = 0.6\,M_\odot$, and $E_{\rm sn,D6} = 1.11\times10^{51}$~erg.

\subsubsection{Morphology of the explosion}

\begin{figure}[ht]
\centering
\includegraphics[width=1\textwidth]{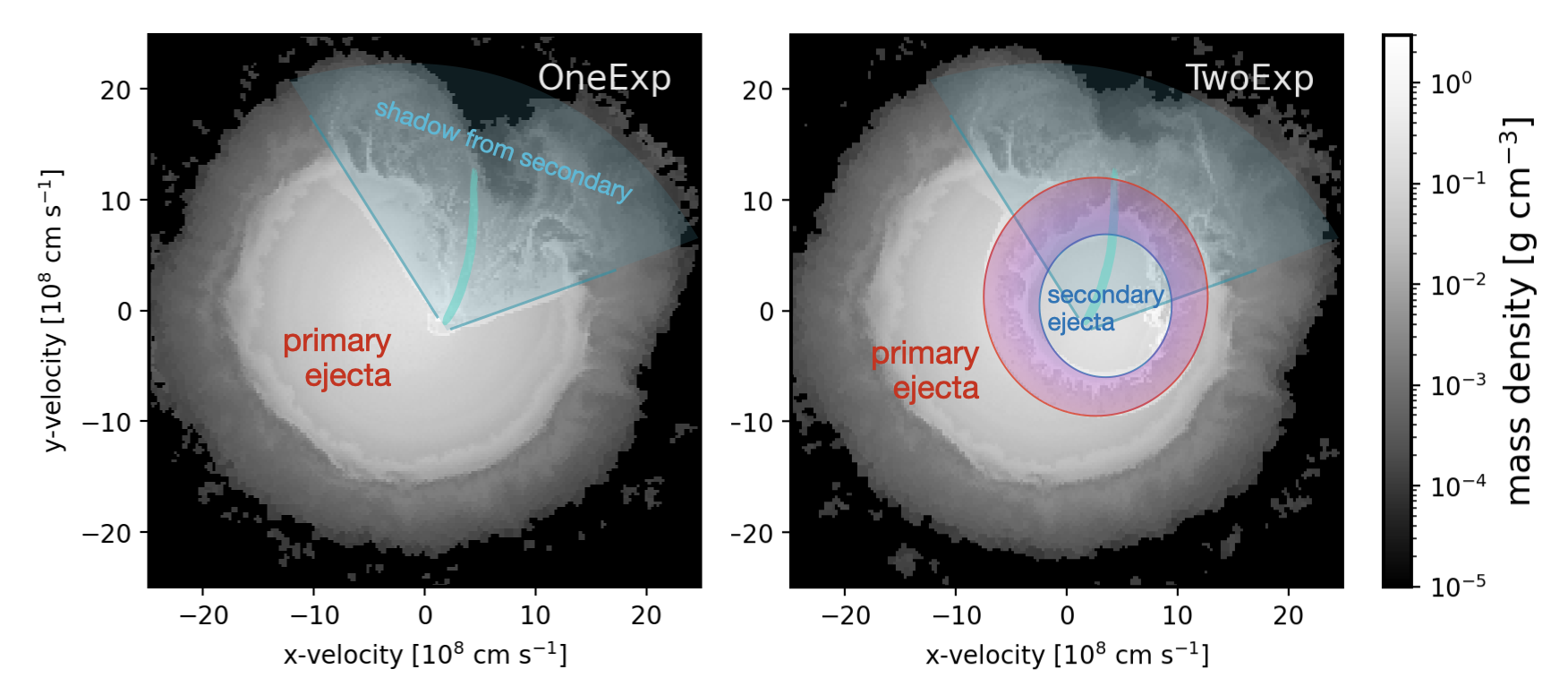}
\caption{Slices of the mass density at the end of the SN simulation (150~s for \OneExp, 133~s for \TwoExp), in the mid-plane along the $z$ axis. We have indicated the ejecta from the primary and from the secondary, as well as the main interaction regions: the angular sector (cyan, a~cone in 3D) is the shadow in the primary ejecta from the presence of the secondary WD, the ring (purple, a shell in 3D) is the double-shock structure formed by the collision of the secondary ejecta into the primary ejecta. The curved segment inside the cone roughly indicates the spiraling motion of the secondary WD.
A~Sketchfab-based 3D interactive version of the \OneExp\ and \TwoExp\ SNe is available online.\footref{Sketchfab} This version shows a set of isocontours of the mass density, rendered as semitransparent surfaces. Contours at 0.004\%, 0.04\%, 0.4\%, 2.3\%, 8\%, 14\%, 20\% of the max were chosen to highlight the extent of the primary ejecta including the shadow and (for \TwoExp) the interaction between the primary and secondary ejecta. 
\label{fig:sketch_OneTwoExp}}
\end{figure}

Slices of the ejecta at the end of the SN simulation for both cases can be seen in \cite{Pakmor2022FateSecondary} (see their Figure~2). We reproduce them with annotations in Figure~\ref{fig:sketch_OneTwoExp}. Interactive 3D models of the ejecta are available online\footnote{\label{Sketchfab}On Sketchfab at \href{https://skfb.ly/puuxH}{https://skfb.ly/puuxH} for \OneExp\ and \href{https://skfb.ly/puuxI}{https://skfb.ly/puuxI} for \TwoExp.} made from isocontours of the mass density, that show the primary and secondary ejecta and their interaction.
One obvious global feature is the cone of low-density ejecta in the direction of the secondary that shadowed them, of opening angle about 45$^{\circ}$ (similar to the model of \citealt{Tanikawa2018Three-DimensionalCompanion}). This happens in both the \OneExp\ and \TwoExp\ cases, although the cone is getting filled in by inner ejecta from the secondary in the \TwoExp\ case. Otherwise, the ejecta are close to spherical symmetry. 
In the \TwoExp\ case, the ejecta from the secondary expand inside the ejecta from the primary, they compress them and quickly get stalled (see the time evolution in Figure~\ref{fig:movie_OneTwoExp} in Appendix~\ref{sec:SN_movie}). By the time the SN ejecta are in homologous expansion, the interaction between the secondary and primary ejecta has well ended -- this is an important point to understand the later evolution. 

Another property is a global velocity shift (w.r.t.\ the rest frame of the binary system) for \OneExp: the ejecta move at $1100\,\mathrm{km}\,\,\mathrm{s}^{-1}$ and the surviving secondary in the opposite direction at $1790\,\mathrm{km}\,\,\mathrm{s}^{-1}$. For \TwoExp\ the net velocity of the centre-of-mass of the ejecta is negligible, as it should be for conservation of momentum.

In terms of composition, the helium ashes are dominated by intermediate-mass elements like Si, S, Ca. Essentially all the radioactive $^{56}$Ni (decayed to Fe within the SNR ages considered) is created from the burnt C-O core of the primary. The mass of $^{56}$Ni is $0.45\,M_\odot$ for \OneExp\ and $0.46\,M_\odot$ for \TwoExp\ (compared with $0.58\,M_\odot$ for the highest resolution D$^6$ of \cite{Tanikawa2018Three-DimensionalCompanion}). For \TwoExp, ashes of the secondary do not contain $^{56}$Ni, but contain intermediate-mass elements because of the much lower initial central density of the secondary. The overall yields for the two cases can be compared in \cite{Pakmor2022FateSecondary}'s Figure~3, and the radial profiles of various elements can be seen in their Figure~4. 

The light curves and spectra were found to be similar for \OneExp\ and \TwoExp. The explosion of the secondary can be completely hidden until well after peak brightness. It is natural to address next whether the secondary ejecta will become better distinguishable in the later phases of the SNR evolution.

\subsection{SNR simulation} 

The output from the SN simulation is used as input for the SNR simulation, conducted with the hydro code Ramses (\citealt{Teyssier2002CosmologicalRAMSES}) in its version developed for SNR studies (\citealt{Ferrand2012Three-dimensionalAcceleration}). Regarding the use of SN simulations as initial conditions, we refer the reader to our previous papers (Ferrand et al. \citeyear{Ferrand2019FromExplosion,Ferrand2021FromModels,Ferrand2022_D6}) for details about our method, including validation and limitations. 

\subsubsection{Initialization and scales}

The SN models are obtained at about 100~s after the explosion (precisely 150~s for \OneExp\ and 133.2~s for \TwoExp), by which time the ejecta are fully homologous, $v\propto r/t$, with a maximum speed $v_{\rm max} \simeq 25,000\,\mathrm{km}\,\,\mathrm{s}^{-1}$. The ejecta profiles, mass density and abundances, are mapped to a $200^3$ grid (of physical size $7.50\times 10^{11}\,\mathrm{cm}$) that is inserted in a $256^3$ grid (to give room for the development of the shocked region). Since most radioactive isotopes have short decay time-scales compared to the SNR ages considered, elemental abundances are computed by fully decaying all radioactive isotopes with half-life times $\le 450$~yr. (The only two radioactive isotopes that may be useful to track individually are $^{56}$Ni and $^{44}$Ti.) When the secondary WD survives, it is not included in the SNR simulation, although its impact on the primary ejecta is accounted for. When the secondary WD explodes, both the secondary and primary ejecta are present. 

Since the very early evolution is self-similar, the start age for the SNR simulation is set to 1~year, with the ejecta profiles rescaled accordingly (we checked that starting at an earlier age, 1~day after the explosion, does not appreciably change the results). We follow the SNR evolution in the ejecta-dominated phase, up to an age of $1500$~yr, which is long enough to reveal the nested ejecta structure of \TwoExp\ (we observed that the SNR evolves slowly after about $1000$~yr). The simulation is performed in a comoving grid, that factors out the global expansion of the SNR (\citealt{Ferrand2019FromExplosion}). As such the physical size of the box explicitly depends on time (and somewhat on the model), over the 1500~yr span it increases by a factor 275 for \OneExp\ and 302 for \TwoExp. 

The expansion of the SNR also depends on the ambient density. In order to separate the effects from the initial and boundary conditions we consider the simplest case of a uniform ISM. As in our previous papers we assume a density of $0.1\,\mathrm{cm}^{-3}$ which, for a typical thermonuclear SN, leads to blast wave dynamics similar to Tycho's SNR at its observed age \citep{Slane2014ModelTycho}, and is consistent with existing density measurements \citep{Williams2013DensityTycho}. We recall that for a young SNR characteristic lengths scale as $M_{\rm ej}^{1/3}$ and $\rho_{\rm ISM}^{-1/3}$.

\subsubsection{Analysis of the morphology}

A~young SNR (regardless of the details of the explosion) is characterized by a shell of shocked matter, bounded by two shocks. The forward shock (FS), or blast wave, travels outwards and shocks the ambient medium, while the reverse shock (RS) travels inwards (in the SNR frame) and shocks the expanding ejecta. The interface between the ejecta and the surrounding matter, known as the contact discontinuity (CD), is unstable; Rayleigh-Taylor (RT) instabilities develop at this interface, producing a distinctive pattern of outward-pointing radial fingers.

The analysis of the simulated SNR morphology is performed as in \cite{Ferrand2019FromExplosion}. At each time step we track in 3D the surfaces of the two shocks (RS and FS) and of the edge of the ejecta (CD). For each wave, we record its radius from the explosion centre, as a function of direction. We record the data on a HEALPix grid, an efficient hierarchical mapping of the sphere with equal-area pixels \citep{Gorski2005HEALPixSphere}. We then expand relative radial variations in spherical harmonics, and compute the angular power spectrum. The simulation was also done with smooth initial conditions, made by averaging the mass density over all angles, in order to disentangle the imprints of the SN explosion from the interaction of the SN ejecta with the ISM. 

\subsubsection{Mock images}

The shocked plasma, between the RS and FS, is heated to X-ray emitting temperatures. The thermal emissivity scales as the density squared, and it depends on the temperature of the electrons, $T_e$, and the ionization age, $n_et$ (where $n_e$ is the electron density and $t$ is the time since the passage of the shock). We first produce maps of the (shocked) mass density squared, which is a commonly used proxy for the thermal X-ray emission, that does not require an assumption about electron heating at the shock or knowledge of the detailed thermodynamic state. As the X-ray emission is optically thin, we sum our estimates of the emissivity over one direction of the grid to obtain maps in projection. The full calculation of the thermal emission, including the detailed non-equilibration ionization balance, will be presented in a forthcoming paper (Fujimaru et al, in prep), together with our previous SNR models. 

\section{Results} 
\label{sec:results}

\begin{figure}[ht]
\centering
\vspace{5mm}
\includegraphics[width=1.00\textwidth]{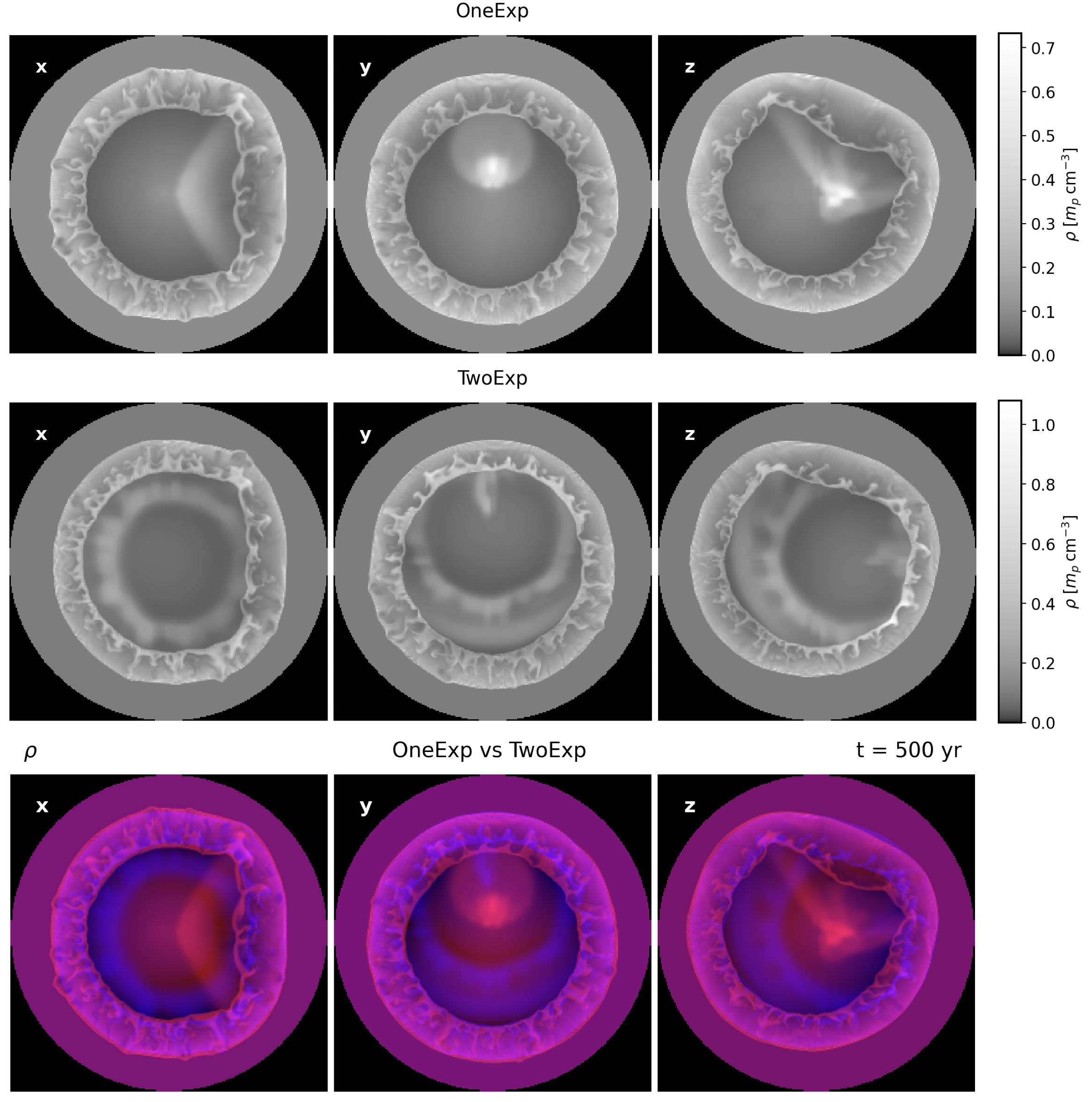}
\caption{Illustration of the making of the comparison maps. The quantity displayed is the mass density, as a slice through the centre of the box, along three different axes, at the age of 500~yr (see more details in section~\ref{sec:res_1} and the time evolution in Figure~\ref{fig:maps_cut_OneTwoExp}). The first two rows show data for \OneExp\ and \TwoExp\ respectively, while the last row shows the composite image that combines the two cases. Data in the first two rows are colourized using a sequential, grayscale colour map. The physical scale is linear but the colour bar has been stretched, by a power 0.33, for better visibility. The last row is a RGB composite of the first two rows, with the \OneExp\ data used as the red channel and the \TwoExp\ data used as the blue channel (the green channel is left at zero), each channel properly normalized in the range $[0,1]$. Where data overlap, the red and blue hues mix into tones of purples. This allows one to see at-a-glance the parts of the SNR that are specific to one case, or that are common to both. (Note that the boundary conditions are spherical, values are not defined in the corners of the box, that are rendered black.)
\label{fig:maps_merging}}
\end{figure}

We now show results from the SNR simulations at selected time snapshots, as well as time-evolution sequences in the form of movies. 
Since the main purpose of this paper is to study the differences between the \OneExp\ and \TwoExp\ cases, we show comparison maps. Specifically, we make RGB composite images with the red channel showing the \OneExp\ data and the blue channel showing the \TwoExp\ data. These images reveal the structure of the SNR, while highlighting parts that are similar (showing up as red + blue = purple) or different (showing up as pure red or pure blue) between the two cases. This is illustrated in Figure~\ref{fig:maps_merging}, that shows the maps separately for the \OneExp\ and \TwoExp\ cases, and the resulting composite image. The shell structure is visible for both cases, with good overall overlap, and small scale variations, especially on the shadow side. The features in the centre, being all red, come from \OneExp, whereas the faint inner ring, being blue, is a feature of \TwoExp. 
The spatial overlaying of maps at different times for the two cases, despite the different physical size, is possible because of the comoving grid. For reference, for the chosen ambient density, the box size is 0.080~pc $|$ 0.080~pc at 1~yr, 5.47~pc $|$ 5.41~pc at 100~yr, 14.5~pc $|$ 14.9~pc at 500~yr, 19.4~pc $|$ 20.7~pc at 1000~yr, 22.0~pc $|$ 24.2~pc at 1500~yr for \OneExp\ $|$ \TwoExp\ respectively.

~\\
\subsection{Global morphology} 
\label{sec:res_1}

\subsubsection{Density slices}

\begin{figure}[ht]
\centering
\vspace{5mm}
\includegraphics[width=0.95\textwidth]{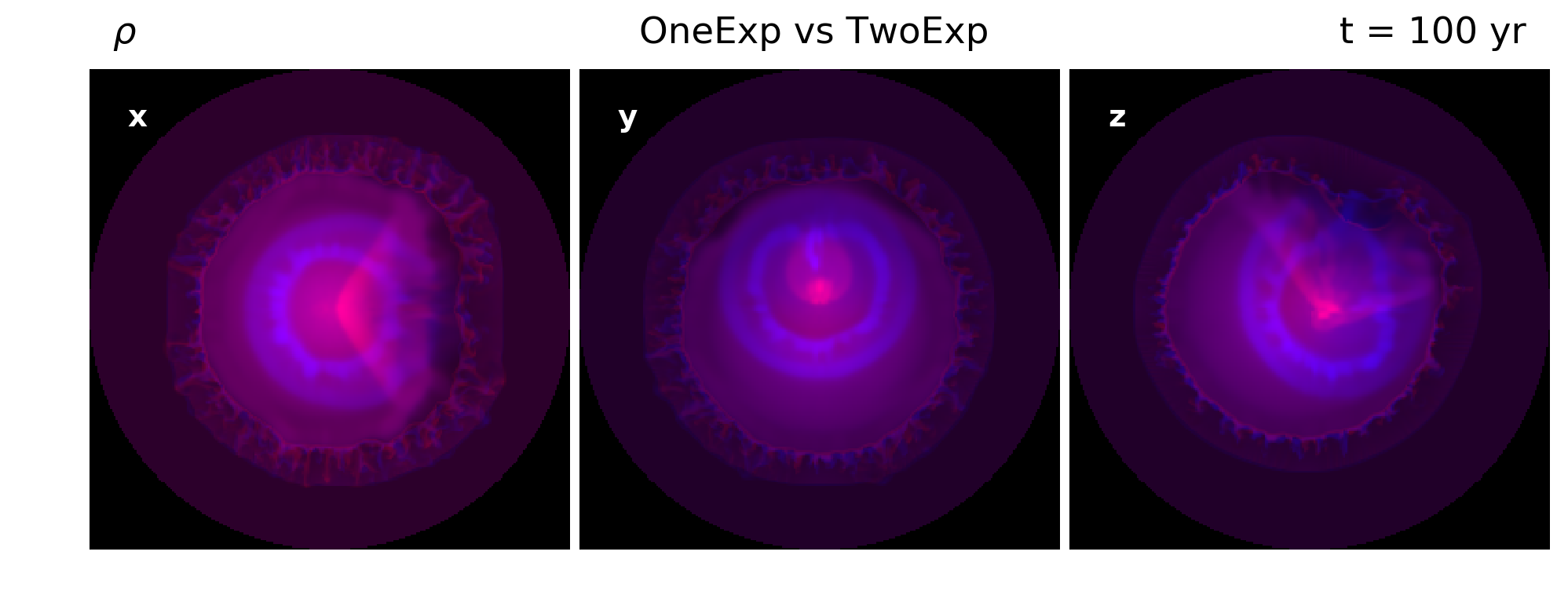}
\includegraphics[width=0.95\textwidth]{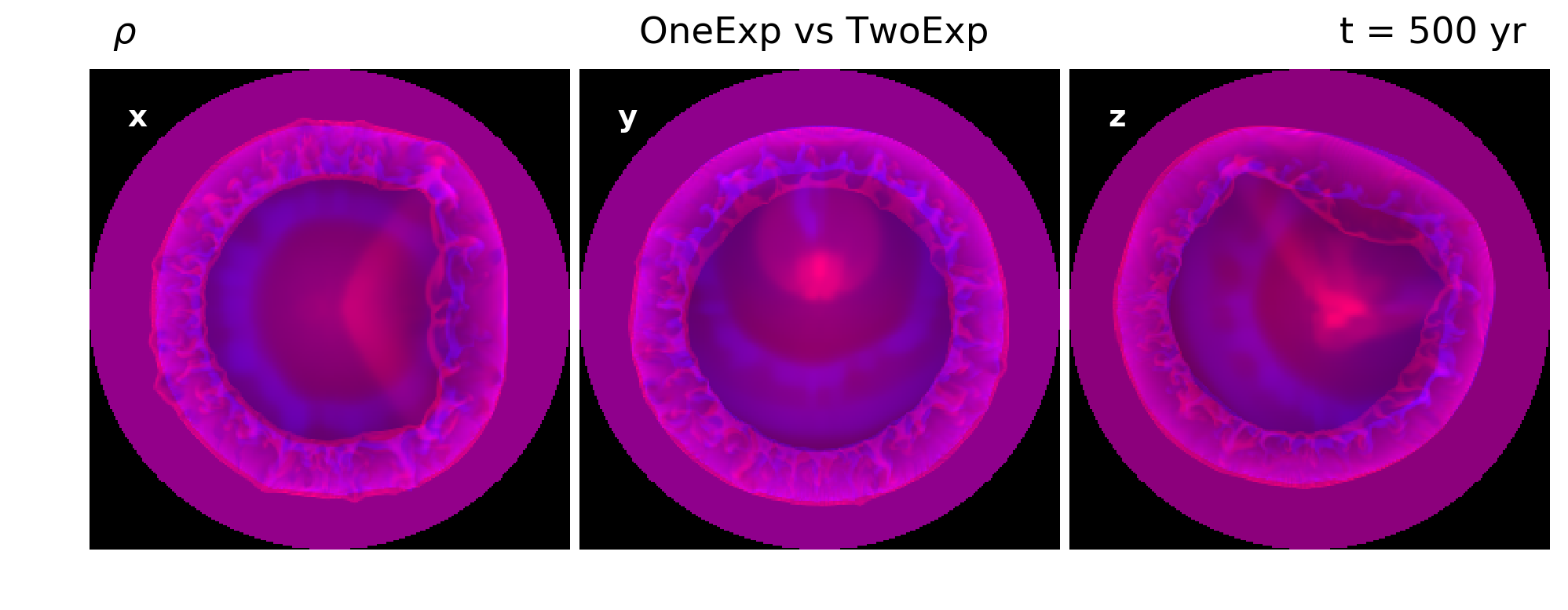}
\includegraphics[width=0.95\textwidth]{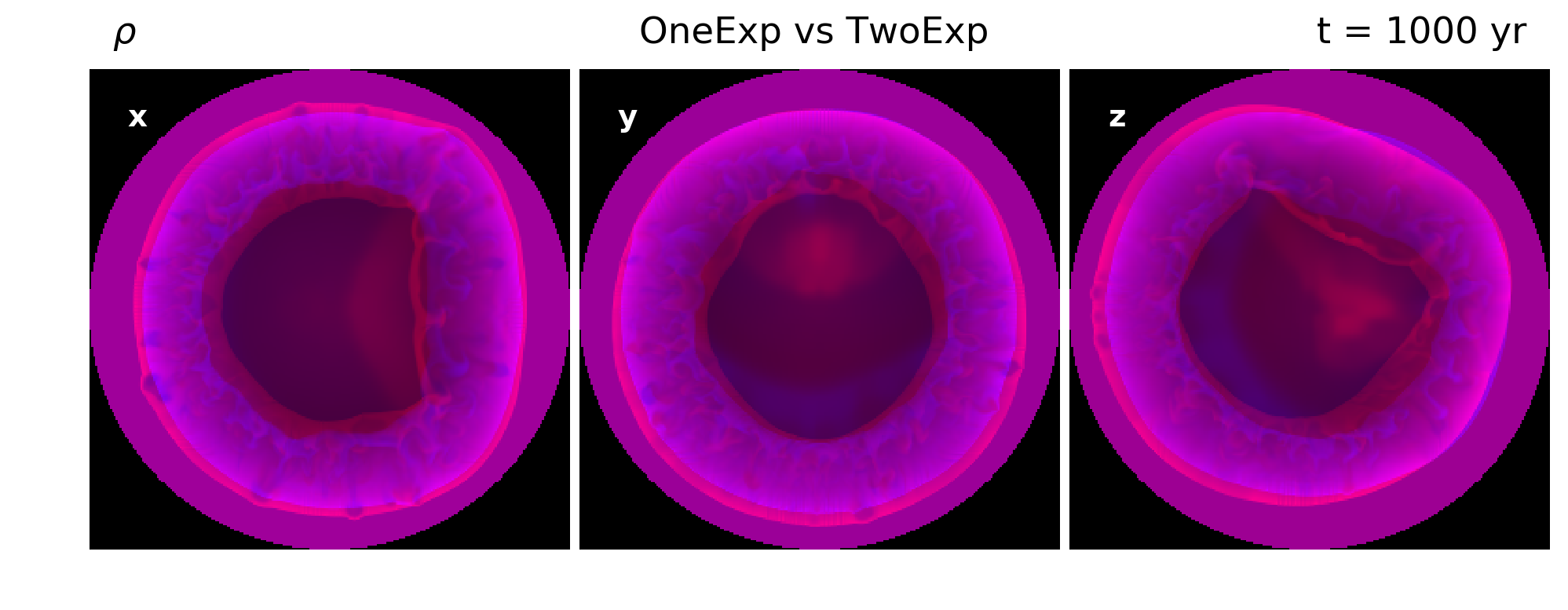}
\caption{Slices of the mass density at several times: 100~yr, 500~yr, 1000~yr after the explosion, along three different axes (principal axes $x$, $y$, $z$, of the simulation box). Each map compares the \OneExp\ and \TwoExp\ cases, with data from \OneExp\ making the red channel and data from \TwoExp\ making the blue channel (purple indicates overlap). Data ranges are normalized for each case and at each time, for each channel a power 0.33 function is applied. The maximum density decreases over time, from 100~yr to 1000~yr by a factor of about 125 for \OneExp\ and 70 for \TwoExp. Each case uses its own comoving grid, for the chosen ambient density, the box size is 5.47~pc $|$ 5.41~pc at 100~yr, 14.5~pc $|$ 14.9~pc at 500~yr, 19.4~pc $|$ 20.7~pc at 1000~yr for \OneExp\ $|$ \TwoExp\ respectively. An animated version of this figure is available in the online journal, showing the evolution from 1~yr to 1500~yr in steps of 1~yr. 
\label{fig:maps_cut_OneTwoExp}}
\end{figure}

In Figure~\ref{fig:maps_cut_OneTwoExp}, we show slices of the mass density at different ages, comparing the \OneExp\ and \TwoExp\ cases. The slicing allows us to see the structure of the SNR. 
The double-shock structure of a young SNR is driven by the hypersonic expansion of the ejecta in the ambient medium.
For the ejecta from the primary WD (for both \OneExp\ and \TwoExp), the ambient medium is the ISM. After just a few years of evolution, both shocks (FS and RS) are clearly visible as sharp discontinuities in the density (as well as in the pressure and velocity). For the ejecta from the secondary WD (for \TwoExp), the ambient medium is the primary ejecta. Already in the initial conditions of \TwoExp\, a small shell is visible embedded inside the ejecta, resulting from the interaction between the secondary and primary ejecta. As we noted in the model section, the secondary explosion is too weak to overtake the primary explosion. The secondary ejecta propagate in the high-density primary ejecta and get stalled well within the first 100~s. Given the homologous expansion was set very early on, the peculiar density structure of the \TwoExp\ ejecta is essentially frozen in time. This includes the pattern of RT fingers, in particular the one pointing inwards that is long lived (visible to the top on the $y$ map and to the right on the $z$ map). 

As can be seen in Figure~\ref{fig:maps_cut_OneTwoExp}, the secondary ejecta stay inside the primary ejecta at all times. The inner ejecta, unshocked primary and secondary, freely expand until they crash on the RS generated by the primary-ISM interaction. This RS processes all of the primary ejecta and then enters into the secondary ejecta. The secondary ejecta are off-centred and they have an asymmetric expansion, since the primary ejecta have different densities in and out of the shadow cone. As a result the secondary ejecta encounter the RS at different times: about 200~yr near the shadow cone, and 600~yr on the opposite side. 

After a few hundred years, the overall morphology of the SNR is not so much affected by either explosion scenario, \OneExp\ or \TwoExp. In both cases, there is effectively a single shell, bounded by a pair of shocks that were generated at the outer edge of the ejecta. The difference is that \TwoExp\ has a non-trivial density structure in the inner ejecta. 
We observe in Figure~\ref{fig:maps_cut_OneTwoExp} that the shocked region is relatively thinner in the \TwoExp\ case when compared to the \OneExp\ scenario. For \TwoExp\ the overall energy in the ejecta is larger, so at a given time the FS is somewhat farther away. However, the secondary ejecta are denser, which slows the inward motion of the RS, resulting in a more compact structure when viewed in the comoving frame.

\subsubsection{Angular structure}

\begin{figure}[ht]
\centering
\includegraphics[width=0.93\textwidth]{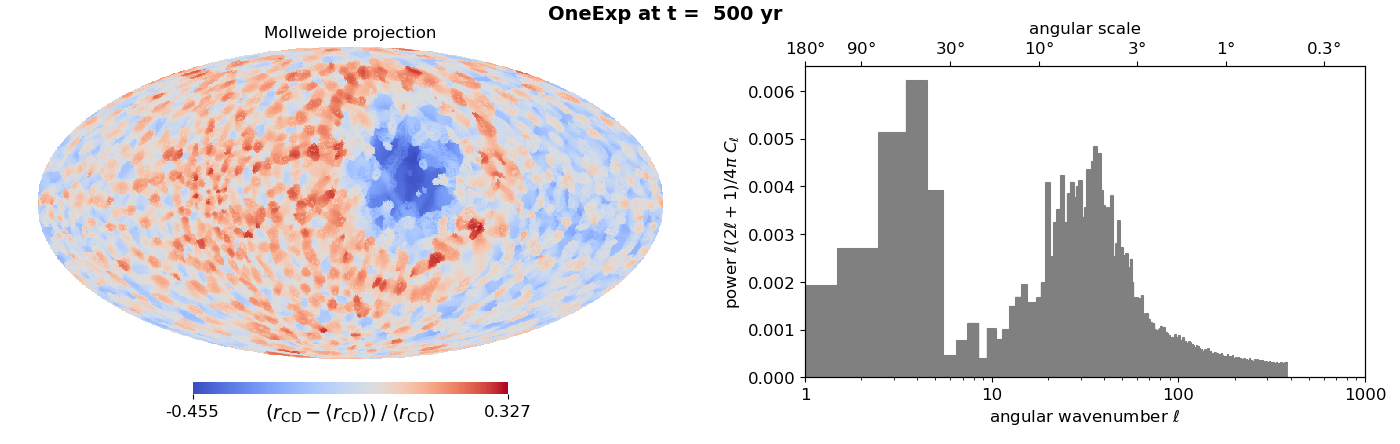}
\includegraphics[width=0.93\textwidth]{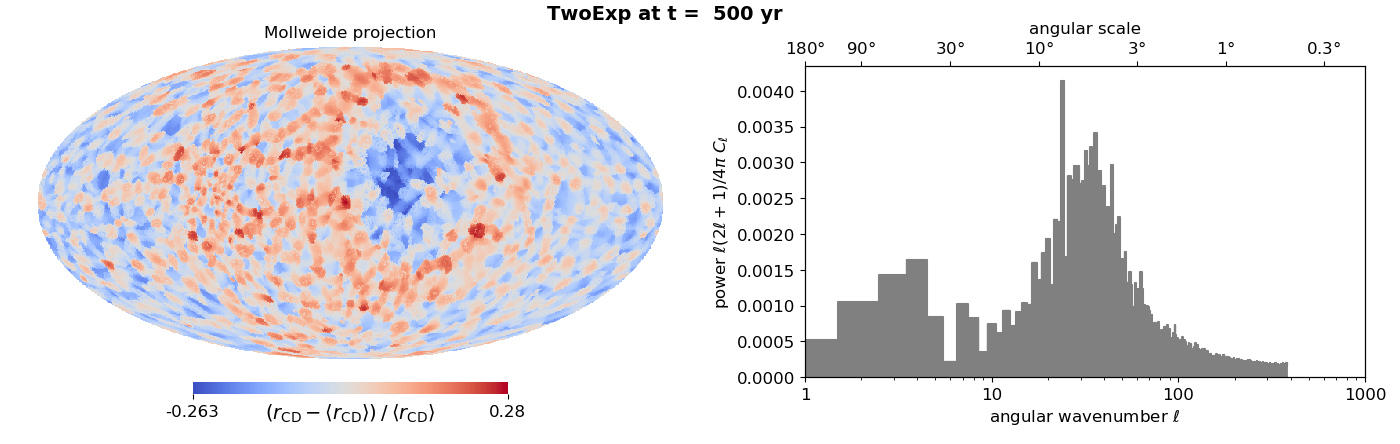}
\caption{Morphology of the contact discontinuity. 
Maps on the left are spherical (Mollweide) projections of the radial variations of the location of the edge of the ejecta. For all times the map is centred on the dipole component at the initial time. 
Spectra on the right result from an expansion in spherical harmonics of these variations. At angular wavenumber~$\ell$, the typical angular scale probed is $\pi/\ell$, and the power $C_{\ell}$ plotted is normalized in such a way that each grayed bin is the contribution of wavenumber~$\ell$ to the total variance of the radial fluctuations.
One time is shown: 500~yr; maps and spectra evolve quickly in the first few hundred years, and slowly after that time.
The two cases \OneExp\ and \TwoExp\ are compared at the top and bottom. 
An animated version of this figure is available in the online journal showing the evolution from 1~yr to 1500~yr in steps of 1~yr. 
Note that data ranges are adjusted for each case and at each time.
\label{fig:healpix_OneTwoExp}}
\end{figure}

\begin{figure}[h]
\centering
\includegraphics[width=0.49\textwidth]{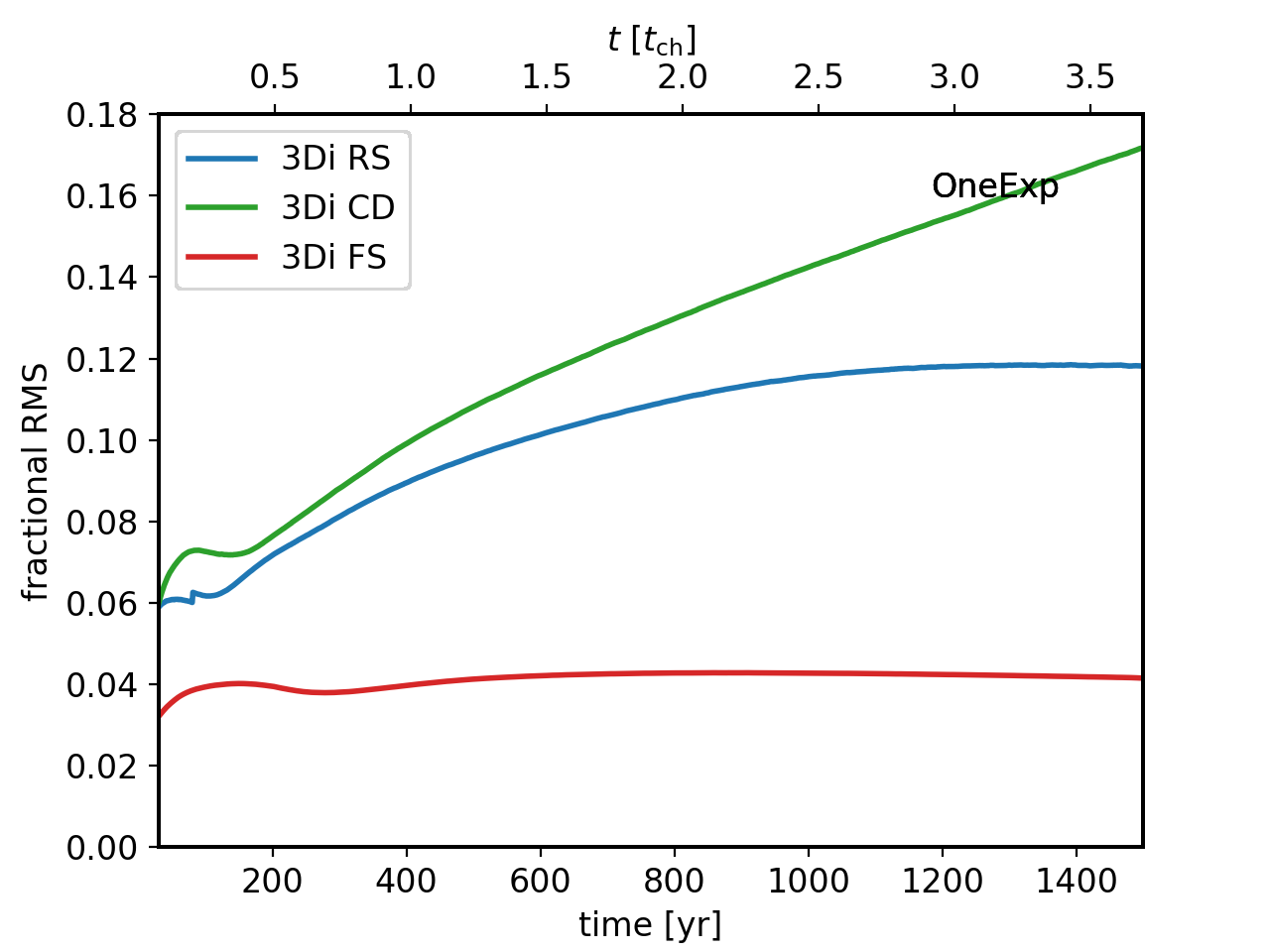}
\includegraphics[width=0.49\textwidth]{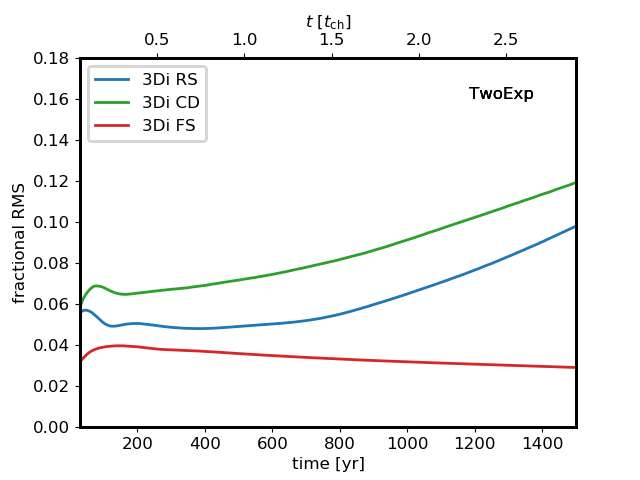}
\caption{Evolution of the angular power as a function of time, for the three waves: forward shock
in red, contact discontinuity in green, and reverse shock in blue. The power is integrated over all angular scales (sum of the gray area on the power spectra), the quantity plotted is its root mean square (RMS). Time is indicated in years and in the characteristic timescale $t_\mathrm{ch} = {r_\mathrm{ch}}/{u_\mathrm{ch}}$ where $r_\mathrm{ch} = \left((3 M_\mathrm{ej})/(4 \pi \rho_\mathrm{ISM})\right)^{1/3}$ and $u_\mathrm{ch} = \left((2E_\mathrm{SN})/(M_\mathrm{ej})\right)^{1/2}$. 
The two cases \OneExp\ and \TwoExp\ are compared on the left and right.
\label{fig:healpix_power_OneTwoExp}}
\end{figure}

In Figure~\ref{fig:healpix_OneTwoExp} we show the radial variations in the location of the outer edge of the SN ejecta (CD with the ambient ISM), over the entire surface of the SNR. The plots on the left are spherical projections, and the plots on the right are angular spectra computed from these. We recall that the angular wavenumber $\ell$ is inversely proportional to the length scale probed. From comparison between simulations made with the actual initial conditions and with a smooth (angularly averaged) version, and from our previous studies, we can tell that large angular scales (small~$\ell$) are dominated by the shape of the SN ejecta that normally regularizes over time, whereas the small angular scales (large~$\ell$) are dominated by the RT instability that always grows over time. This bimodal distribution is clear on the angular spectra. There is more power at large scales for \OneExp\ than for \TwoExp, because the shadow structure is clearer for the former case. Even though some differences can be seen between \OneExp\ and \TwoExp, the overall morphology of the outer edge of the ejecta is the same. The shape of the RS, which is very close to the CD, follows a similar pattern (not shown). The shape of the FS is even less affected by what is happening inside the ejecta (not shown). Given our assumption of a uniform ISM, the blast wave shows much less structure than the other waves, although it can be locally deformed by the tips of the RT fingers (see Figure~\ref{fig:maps_cut_OneTwoExp}). 

In Figure~\ref{fig:healpix_power_OneTwoExp} we show the time evolution of the total angular power for the three waves, for both the \OneExp\ and \TwoExp\ cases. The CD has the most complex structure and its angular power keeps growing in time. The concavity of the curve is different for \OneExp\ vs. \TwoExp, for \OneExp\ the evolution is faster at the beginning. The FS has a simple and quite steady shape. The RS has an intermediate evolution: in the \OneExp\ case, as the RS and CD get more separated structures no longer grow at the RS, whereas in the \TwoExp\ case, as the RS is probing the outer then inner ejecta its shape keeps evolving. Notably, contrary to our previous simulations with a grid of SN models without a companion (\citealt{Ferrand2021FromModels}), the power at the CD rises in time at both small scales and large scales. At small scales this is caused by the RT instability, at large scales it is the imprint of the shadow from the secondary~WD.

Because of the lower density of the ejecta in the shadow region, the CD and the RS are deformed, they assume a flat rather than curved shape within the cone. Conversely the edge of the cone has its density enhanced. When looking at the slices in Figure~\ref{fig:maps_cut_OneTwoExp}, the geometry depends on the direction of observation: the effect is most obvious along~$z$ (where the cone is seen sideways), also clear along~$x$, not visible along~$y$ (where the cone is seen almost along its axis). The circular outline of the cone is clearly visible on the surface of the ejecta in Figure~\ref{fig:healpix_OneTwoExp}. 

As we have seen so far, distinguishing the \OneExp\ and \TwoExp\ cases requires being able to look inside the SNR, which is possible to some extent using the thermal X-ray emission from the shocked ejecta. This emission is optically thin, and so allows one to see the entire SNR and from any direction.

\subsection{Maps in projection} 
\label{sec:res_2}

The X-ray emission from the SNR comes from the high-temperature shocked matter, both the shocked ejecta (located between the RS and the CD) and the shocked ISM (located between the CD and the FS). For both cases, \OneExp\ and \TwoExp, this structure refers to the main, outer shell. Indeed for \TwoExp\ the inner secondary ejecta are quickly diluted by the expansion, their pressure is negligible at all times like the rest of the expanding ejecta. So we do not expect X-ray emission from the inner mini-shell that is visible in the early hydrodynamic slices. It is the RS from the interaction with the ISM that shock-heats the ejecta to X-ray-emitting temperatures. However, for the \TwoExp\ case the presence of the embedded secondary ejecta creates, compared with the \OneExp\ case, a more complex structure of the density and composition of material probed by the RS in the SNR phase.

\begin{figure}[ht]
\centering
\vspace{5mm}
\includegraphics[width=0.95\textwidth]{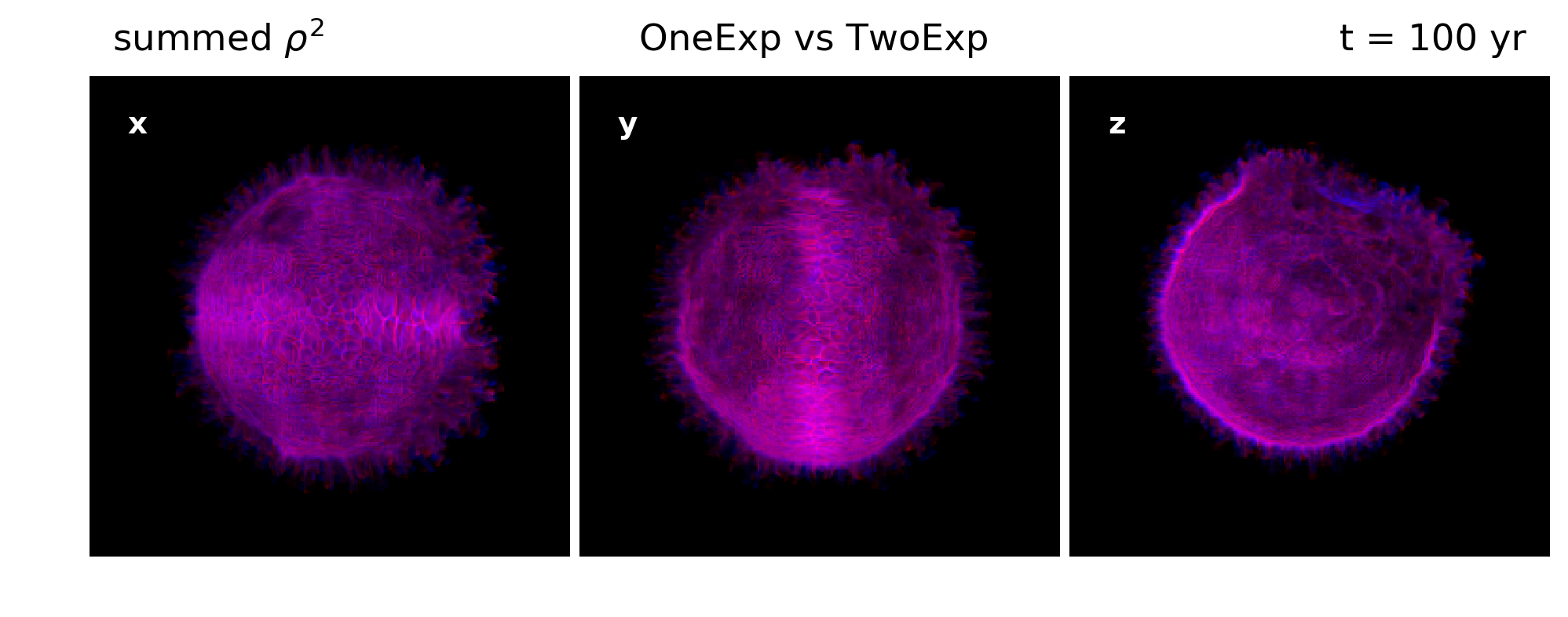}
\includegraphics[width=0.95\textwidth]{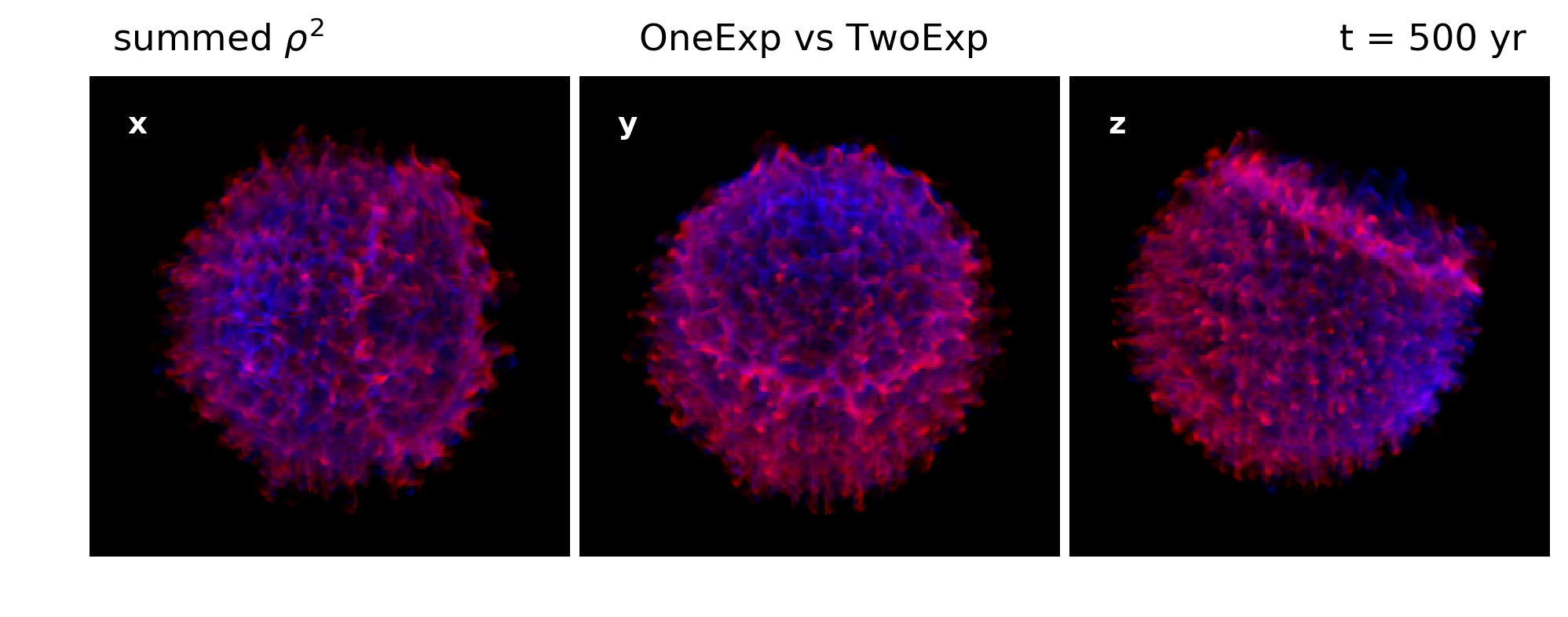}
\includegraphics[width=0.95\textwidth]{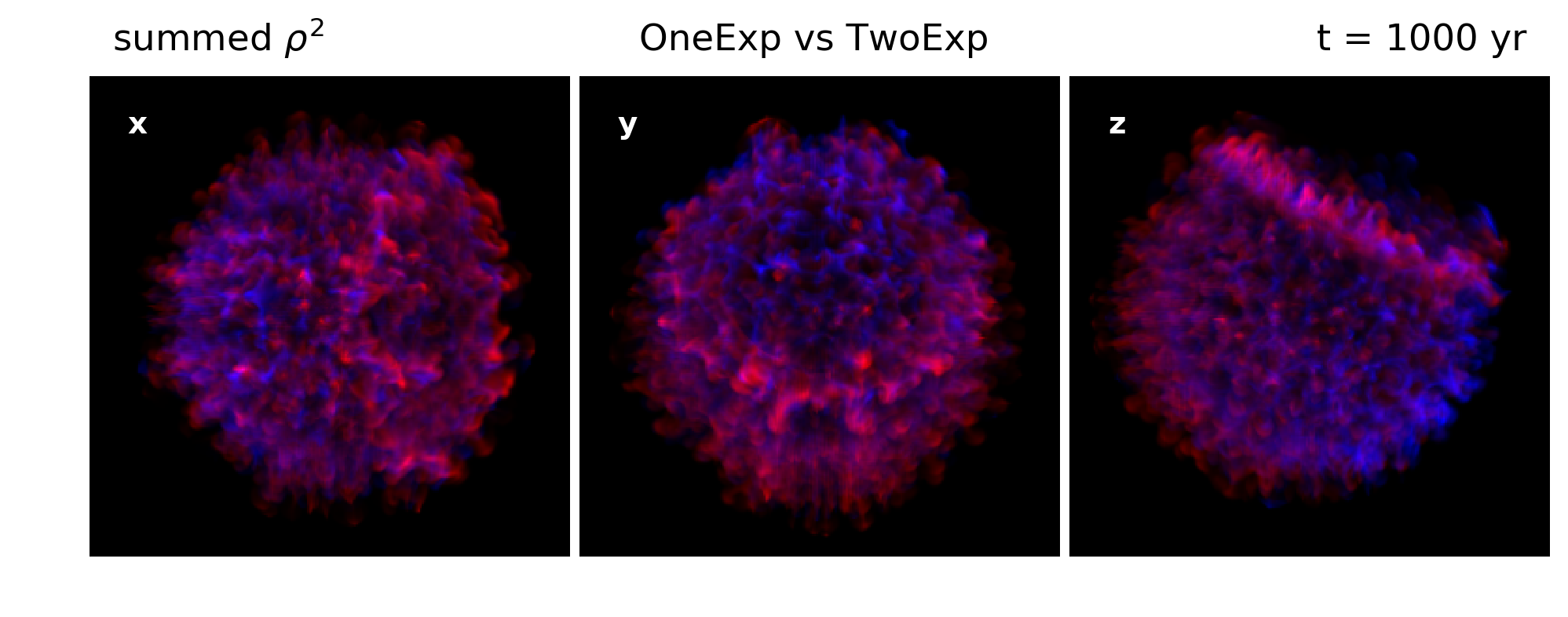}
\caption{Same as Figure~\ref{fig:maps_cut_OneTwoExp}, for the projection of the mass density squared of the shocked ejecta. The \OneExp\ case is the red channel and the \TwoExp\ case the blue channel.
An animated version of this figure is available in the online journal, showing the evolution from 1~yr to 1500~yr in steps of 1 yr. Data ranges are normalized for each case and at each time, for each channel the scale is linear. The maximum value decreases over time, from 100~yr to 1000~yr by a factor of about 1350 for \OneExp\ and 270 for \TwoExp. 
\label{fig:maps_prj_OneTwoExp}}
\vspace{30mm}
\end{figure}

\subsubsection{Overall emission}

In Figure~\ref{fig:maps_prj_OneTwoExp} we show a commonly used proxy for the thermal emission: the density squared (of the shocked matter) summed along the line of sight, comparing the \OneExp\ and \TwoExp\ cases. The small-scale granularity on these maps in projection is caused by the RT instability, whereas at large scales an imprint from the SN may remain. 
As stated previously, the \OneExp\ case would be almost spherical, except for the shadow from the secondary, the impact of which is visible from a few hundred years to more than a thousand years after the explosion. The inside of the shadow cone, being of lower density, appears as missing emission, while the edge of the cone, being of enhanced density, appears as brighter emission. In projection this effect is always visible, although in a way that depends on the direction of observation. When looking mostly along the axis of the cone (along~$y$, plots in the middle), one can see a large circled hole; when looking at some angle (along~$x$, plots on the left), one can see an ellipse; and when looking on the side of the cone (along~$z$, plots on the right), one can see a bright bar. 

Another feature that is visible on the maps, for both the \OneExp\ and \TwoExp\ cases, is a localized enhancement in brightness around 100~yr after the explosion. It appears as a horizontal and vertical bar, respectively, on the $x$ and $y$ maps, and as a ring on the $z$~map (incomplete because of the shadow in the top right corner). So it is shaped like a belt in the $x$-$y$ plane, which is the plane of orbital motion. Looking at the density slices in Figure~\ref{fig:maps_cut_OneTwoExp}, this corresponds to the RS entering a plateau in the density of the primary's ejecta, at about two thirds of their maximal radius. This step is visible on the averaged profile of Figure~4 in \cite{Pakmor2022FateSecondary} at around two-thirds of the radial axis. It appears to be where the (primary) ejecta transition from being He shell ashes to being CO core ashes. 

For the \TwoExp\ case, an additional feature that comes from the exploded secondary is visible. Where the secondary ejecta collide with the main shell (from the primary-ISM interaction), which for our conditions starts happening after a couple hundred years, there is a marked over-density and thus a bump in emission. This is particularly visible at the top of the $y$~maps and at the bottom right of the $z$~maps, this happens in the centre on the $x$~map. See Figure~\ref{fig:maps_cut_OneTwoExp} for a comparison with the density slices and to identify the progressive collision of the secondary ejecta with the primary~RS. 

\subsubsection{Emission by elements}

\begin{figure}[ht]
\centering
\vspace{5mm}
\includegraphics[width=0.95\textwidth]{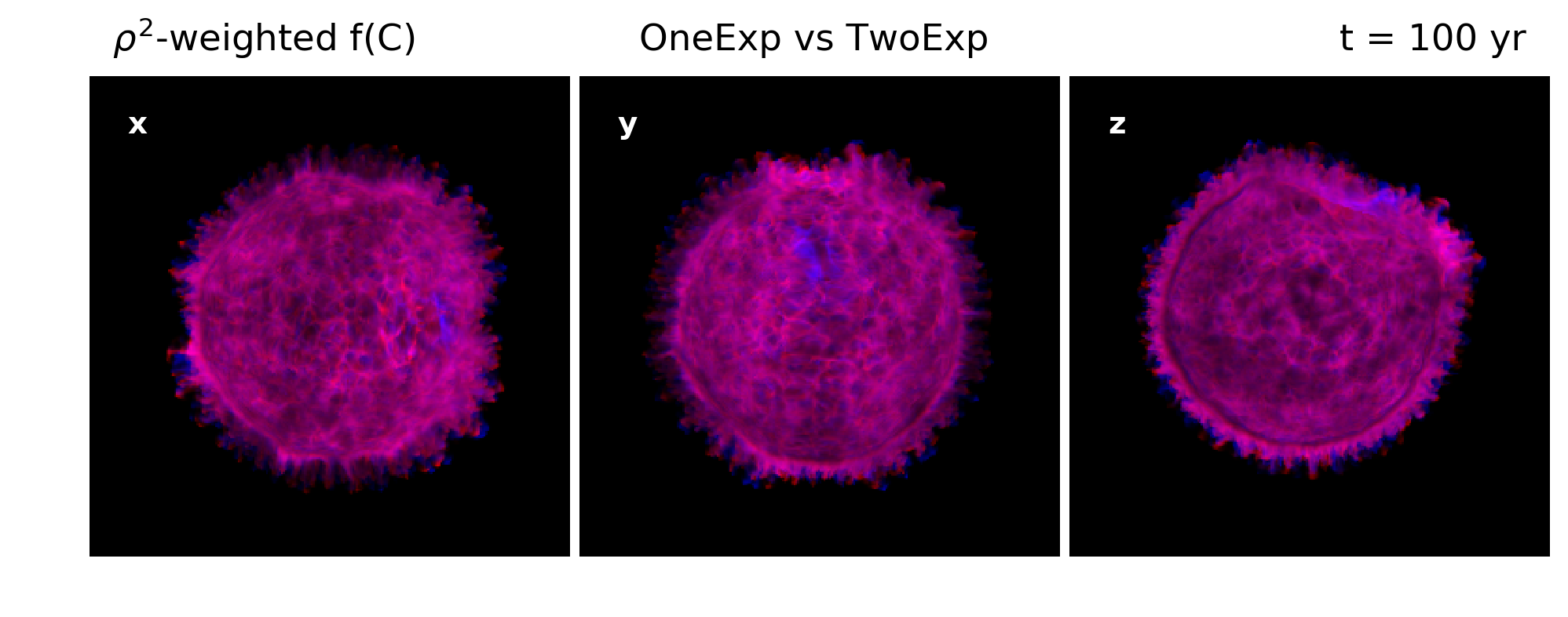}
\includegraphics[width=0.95\textwidth]{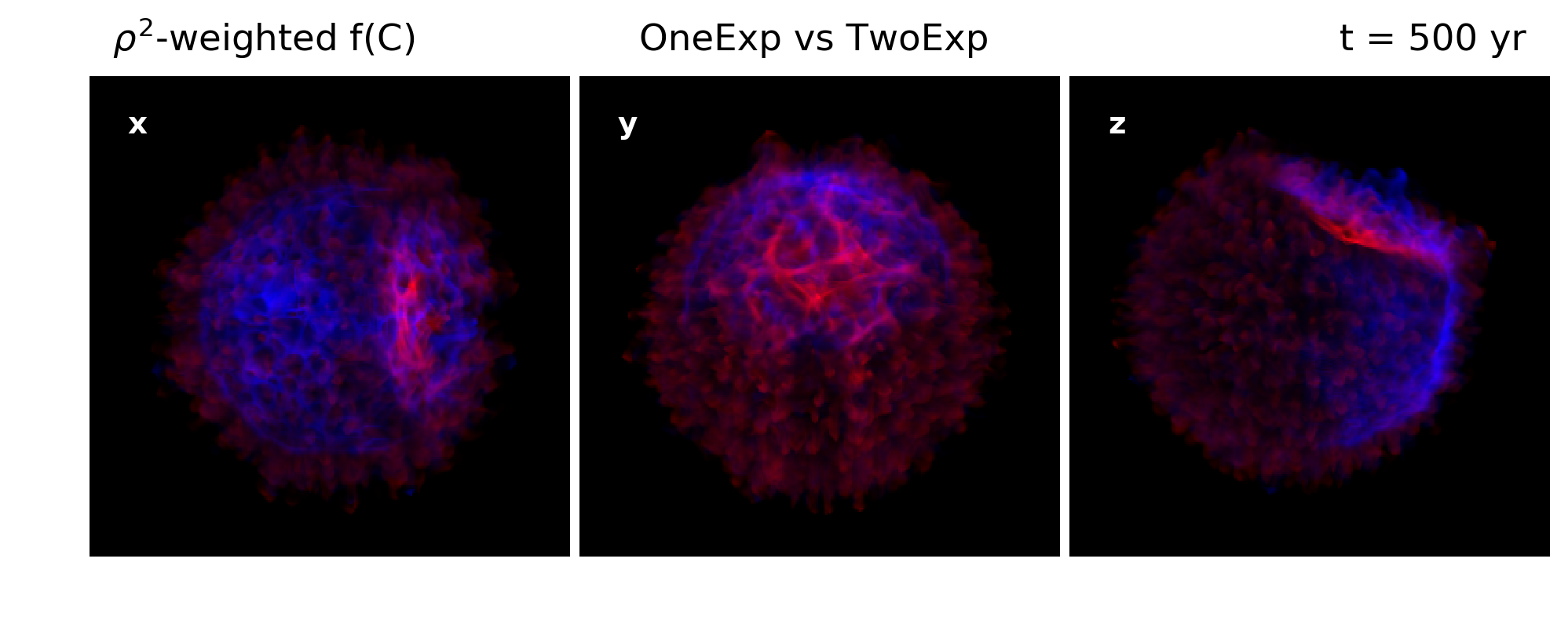}
\includegraphics[width=0.95\textwidth]{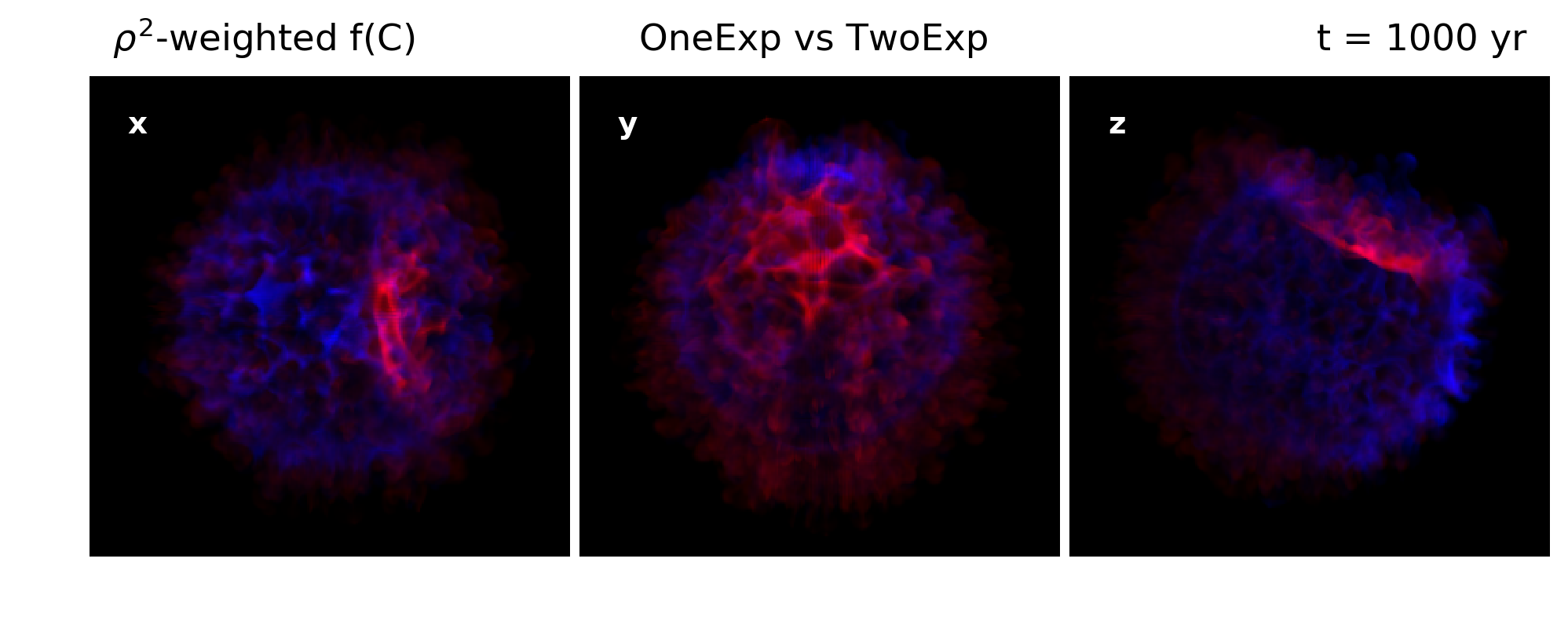}
\caption{Same as Figure~\ref{fig:maps_prj_OneTwoExp}, for the projection of the mass fraction of carbon $f(C)$ in the shocked ejecta, weighted by the density squared. The \OneExp\ case is the red channel and the \TwoExp\ case the blue channel.
An animated version of this figure is available in the online journal, showing the evolution from 1~yr to 1500~yr in steps of 1 yr. 
\label{fig:maps_prj_C_OneTwoExp}}
\vspace{30mm}
\end{figure}

\begin{figure}[ht]
\centering
\vspace{5mm}
\includegraphics[width=0.95\textwidth]{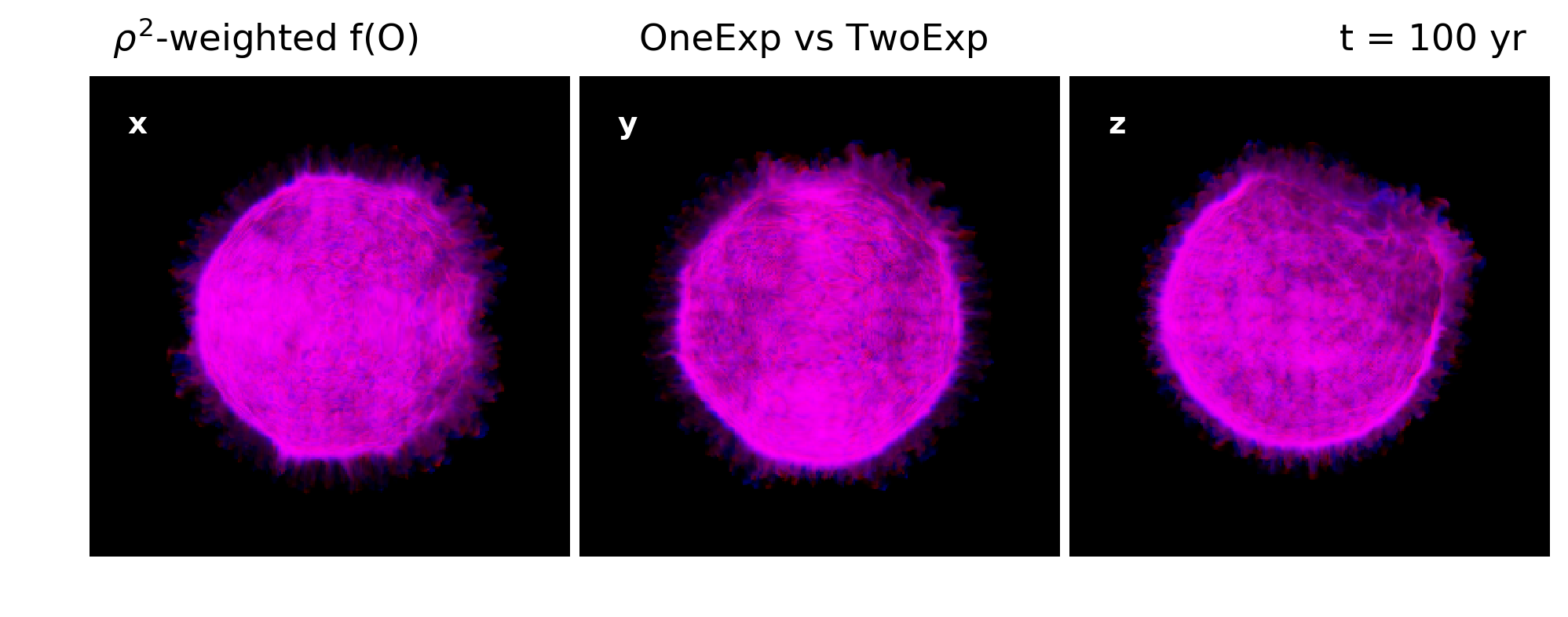}
\includegraphics[width=0.95\textwidth]{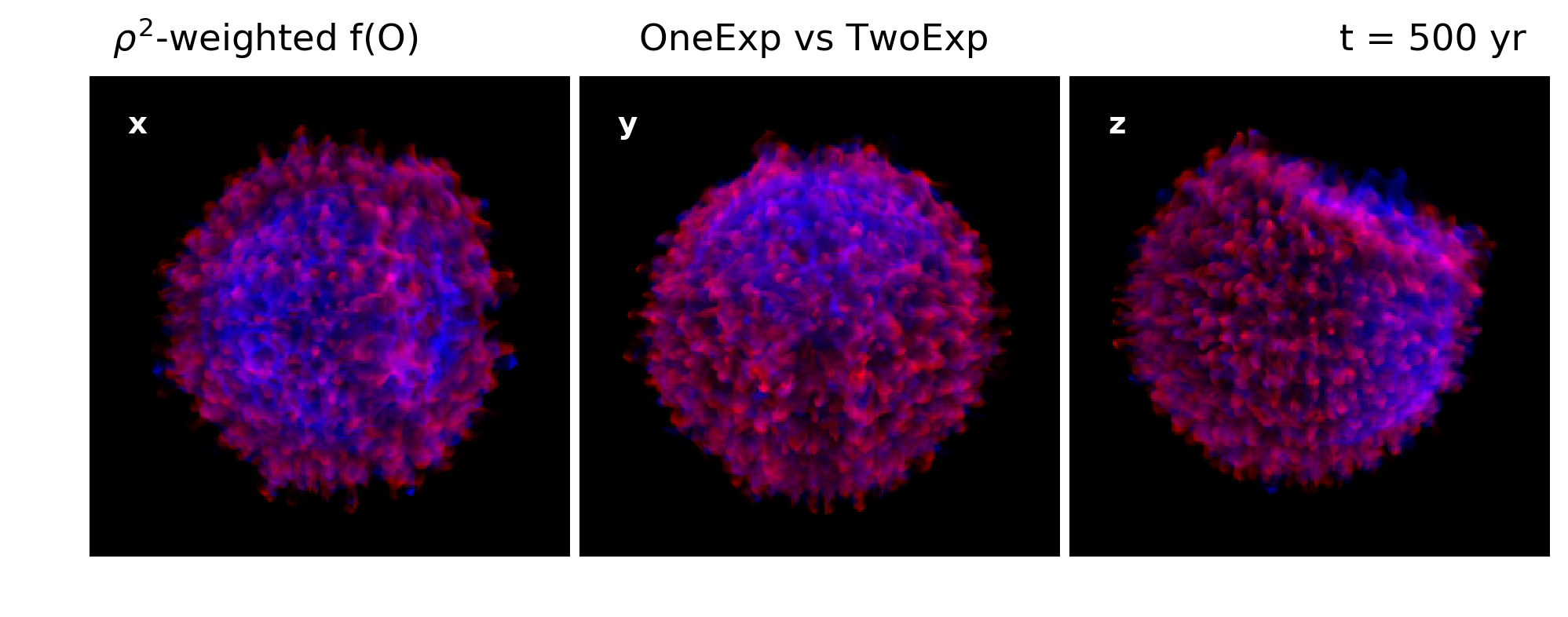}
\includegraphics[width=0.95\textwidth]{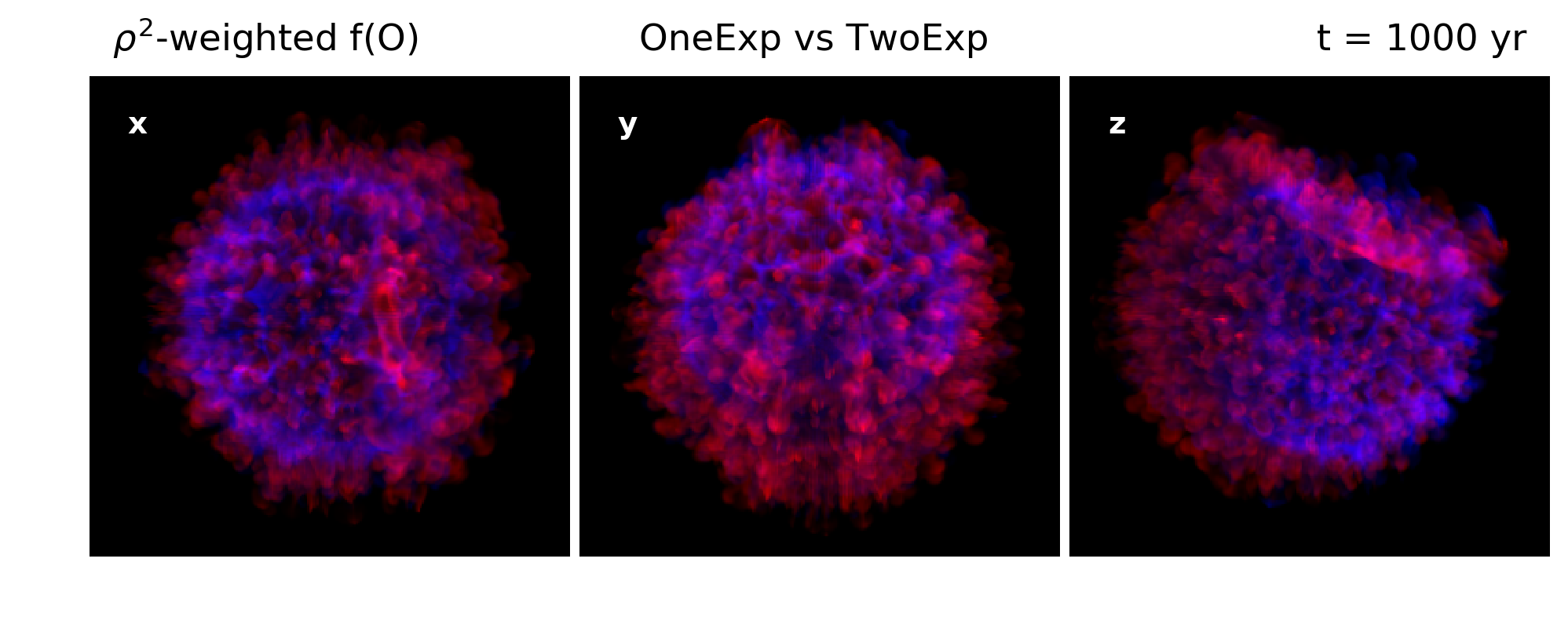}
\caption{Same as Figure~\ref{fig:maps_prj_OneTwoExp}, for the projection of the mass fraction of oxygen $f(O)$ in the shocked ejecta, weighted by the density squared. The \OneExp\ case is the red channel and the \TwoExp\ case the blue channel.
An animated version of this figure is available in the online journal, showing the evolution from 1~yr to 1500~yr in steps of 1 yr. 
\label{fig:maps_prj_O_OneTwoExp}}
\vspace{30mm}
\end{figure}

\begin{figure}[ht]
\centering
\vspace{5mm}
\includegraphics[width=0.95\textwidth]{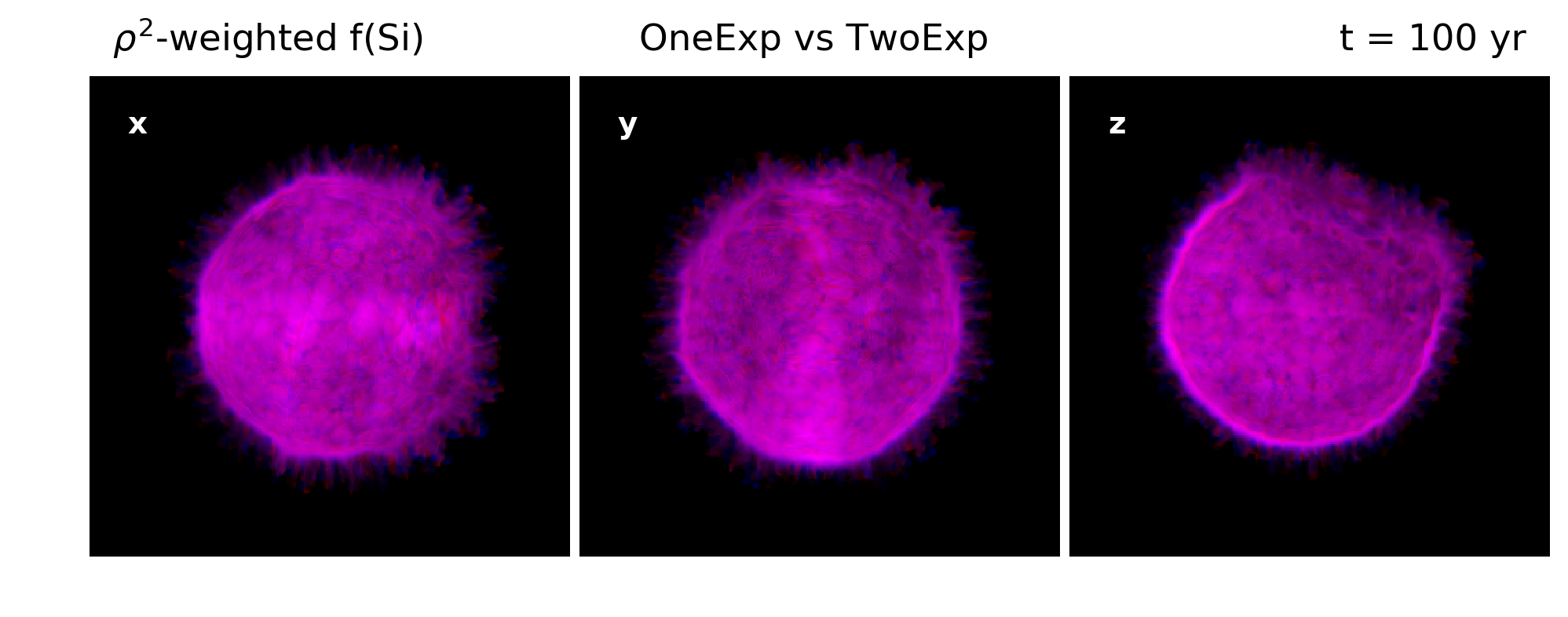}
\includegraphics[width=0.95\textwidth]{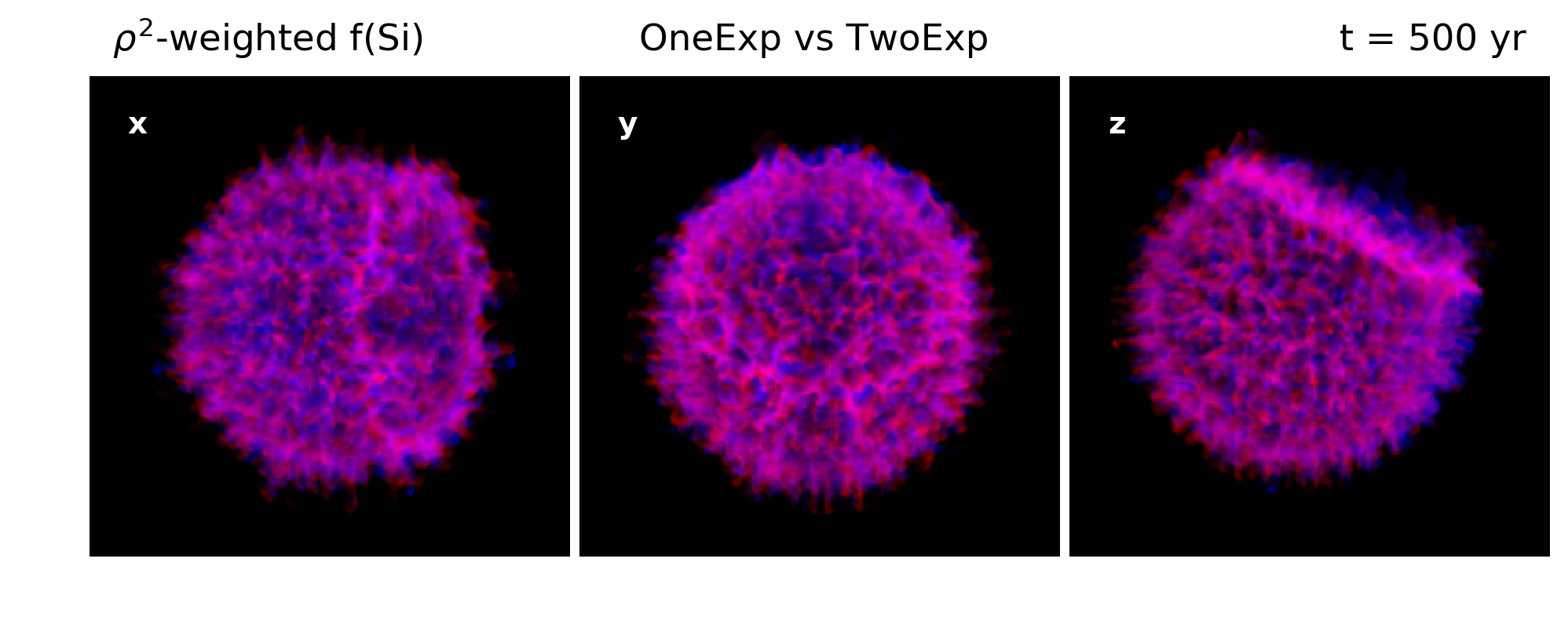}
\includegraphics[width=0.95\textwidth]{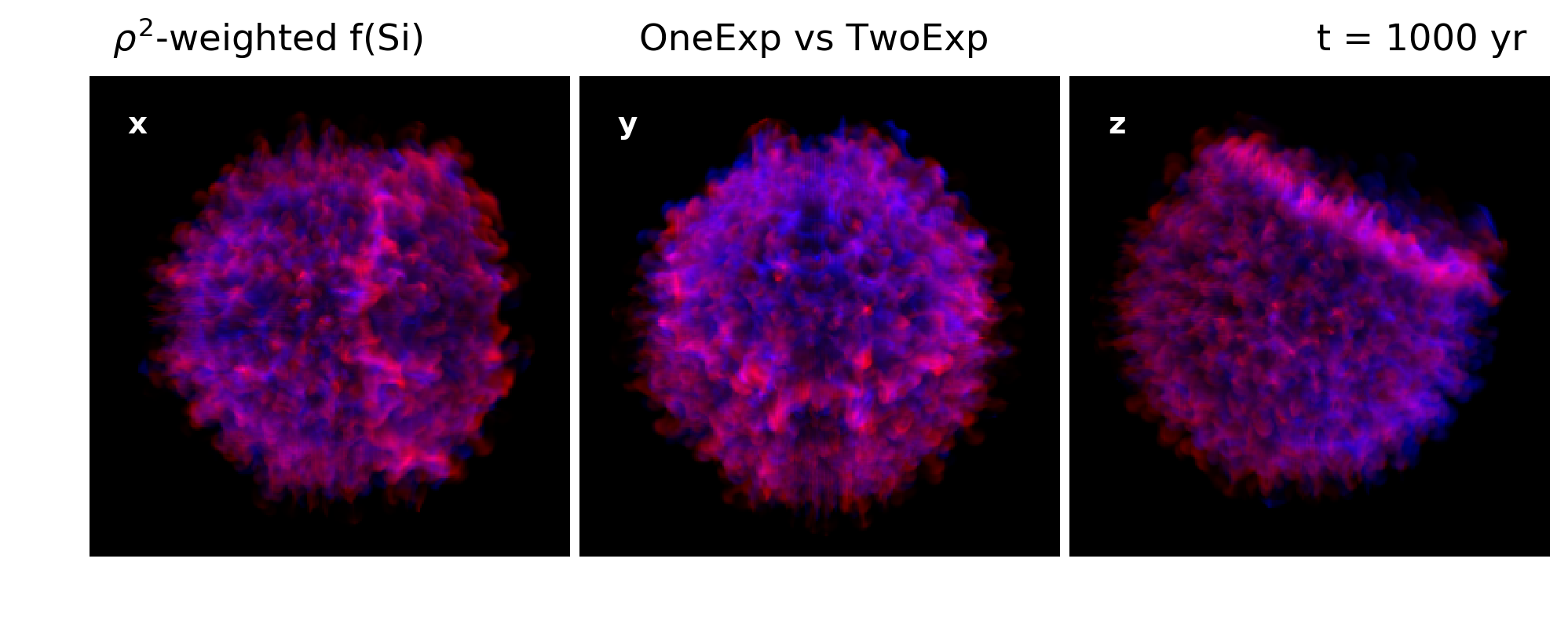}
\caption{Same as Figure~\ref{fig:maps_prj_OneTwoExp}, for the projection of the mass fraction of silicon f(Si) in the shocked ejecta, weighted by the density squared. The \OneExp\ case is the red channel and the \TwoExp\ case the blue channel.
An animated version of this figure is available in the online journal, showing the evolution from 1~yr to 1500~yr in steps of 1 yr. 
\label{fig:maps_prj_Si_OneTwoExp}}
\vspace{30mm}
\end{figure}

\begin{figure}[ht]
\centering
\vspace{5mm}
\includegraphics[width=0.95\textwidth]{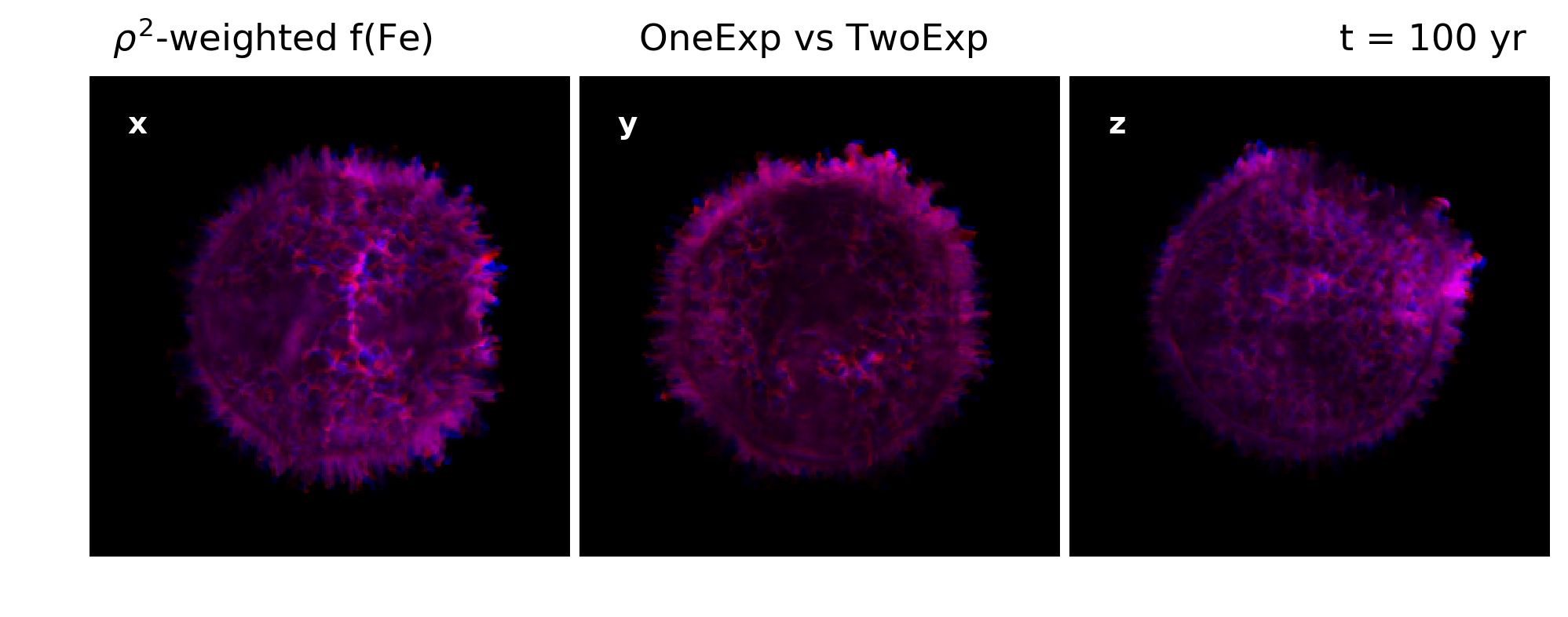}
\includegraphics[width=0.95\textwidth]{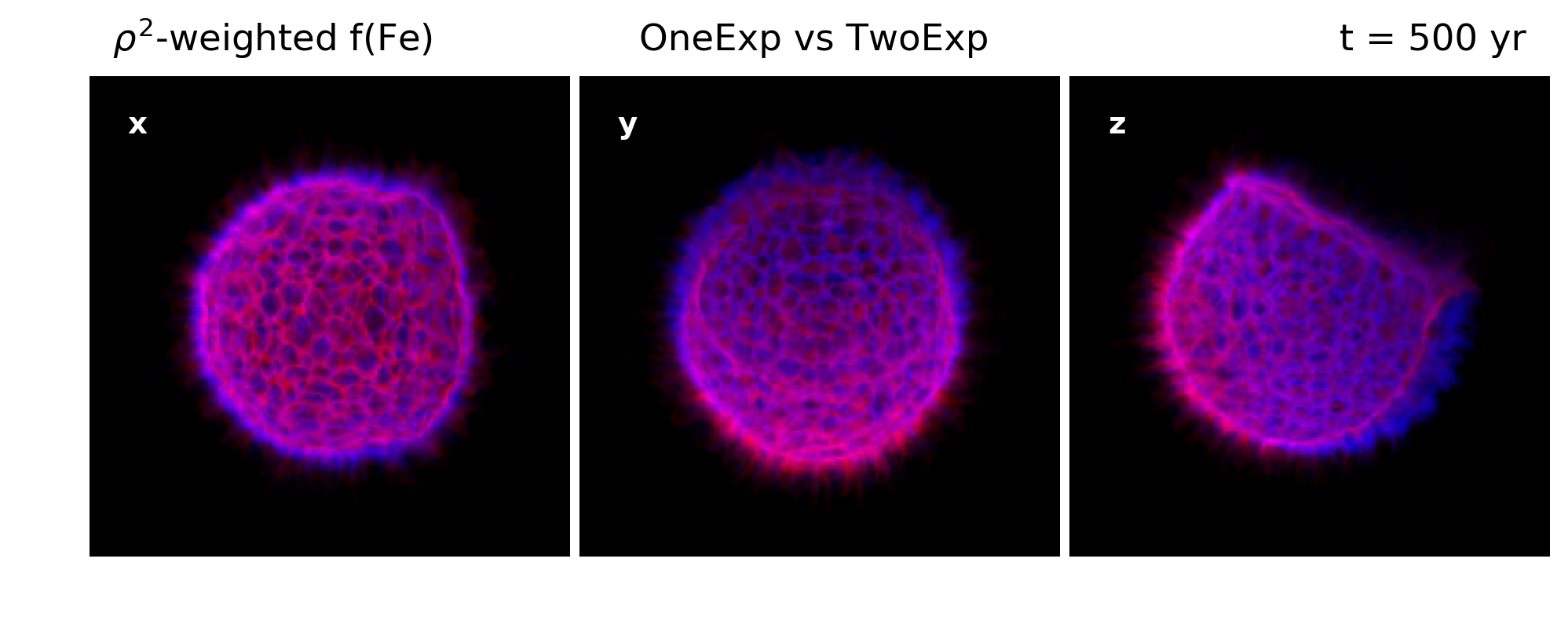}
\includegraphics[width=0.95\textwidth]{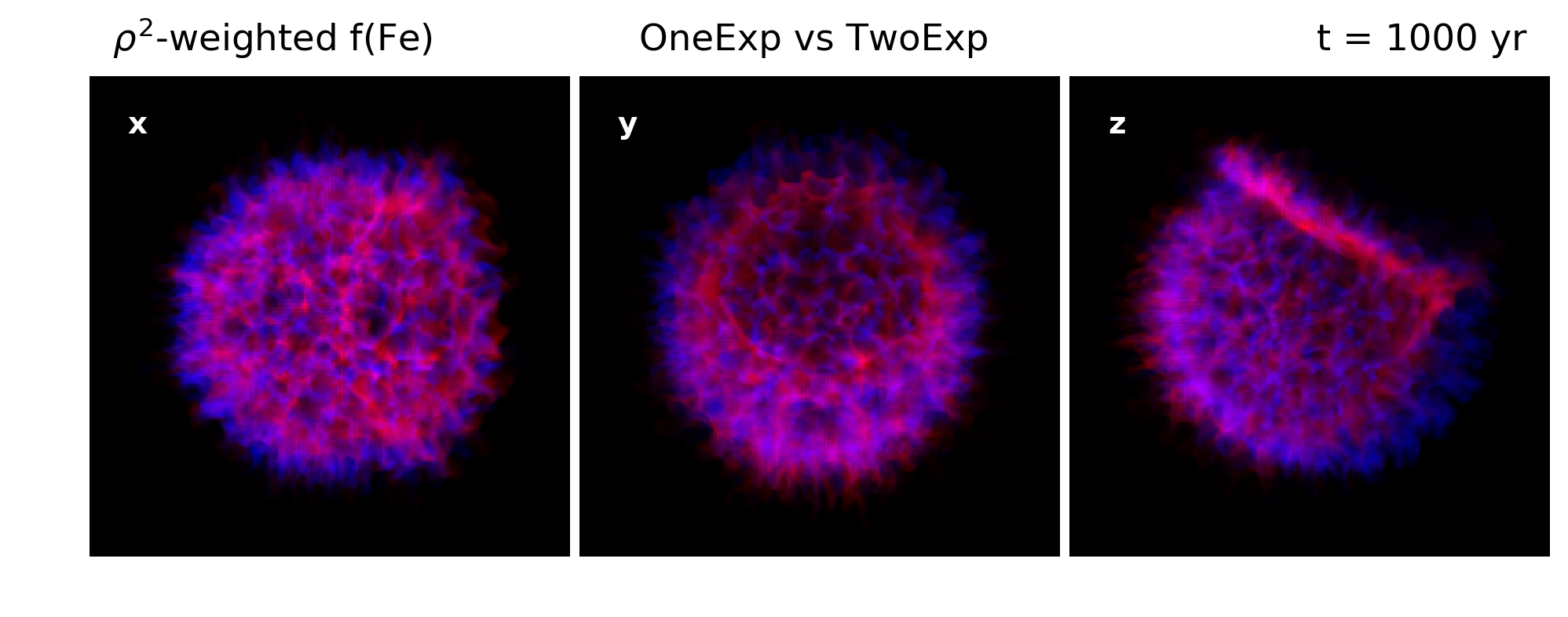}
\caption{Same as Figure~\ref{fig:maps_prj_OneTwoExp}, for the projection of the mass fraction of iron f(Fe) in the shocked ejecta, weighted by the density squared. The \OneExp\ case is the red channel and the \TwoExp\ case the blue channel.
An animated version of this figure is available in the online journal, showing the evolution from 1~yr to 1500~yr in steps of 1 yr. 
\label{fig:maps_prj_Fe_OneTwoExp}}
\vspace{30mm}
\end{figure}

We go one step further by considering the composition of the plasma. We show the abundance of select elements: in Figure~\ref{fig:maps_prj_C_OneTwoExp} for carbon, Figure~\ref{fig:maps_prj_O_OneTwoExp} for oxygen, Figure~\ref{fig:maps_prj_Si_OneTwoExp} for silicon, Figure~\ref{fig:maps_prj_Fe_OneTwoExp} for iron, as before comparing the \OneExp\ and \TwoExp\ cases. As in the previous maps, the elemental abundance (mass fraction) is summed along the line of sight and weighted by the density squared, as a proxy for the X-ray emissivity (in the energy bands where these elements have emission lines). These maps show where elements may be detected, without getting into the details of the ionization state of the plasma. 

For both \OneExp\ and \TwoExp\ unburnt carbon is visible chiefly at the edge of the shadow cone, but for \TwoExp\ there is an excess at the collision sites that becomes dominant after 500 years (we recall that this happens in the middle along~$x$, on the top along~$y$, on the lower right side along~$z$). 
Oxygen (unburnt or carbon ashes) is well spread over the remnant, with a strong excess at the collision sites for \TwoExp. The transient over-dense belt previously noted at 100~yr is clearly visible for~O, also for~Si, afterwards it appears as a darker region. 
Silicon (as well as some other intermediate mass elements) is visible all over, and enhanced at the rim of the shadow, for both \OneExp\ and \TwoExp. In time an excess appears at the collision sites for \TwoExp. 
Iron (from nickel decay) is located more inside and is asymmetrically limb-brightened at 500 years, in time it transitions from the feet to the tips of the RT fingers. Its distribution is slightly different between the \OneExp\ and \TwoExp\ cases. For the \TwoExp\ case, the collision is visible only transiently because there is no more iron in the inner ejecta, once the RS has processed all the primary ejecta. For this case, at 1000~yr iron is mainly located on the side opposite to the secondary ejecta.

\section{Discussion} 
\label{sec:discussion}

We refer the reader to our previous papers, the latest in the series being \cite{Ferrand2022_D6} (see section~4.1), for the possible limitations of our physical modeling. We discuss here how our findings compare with previous SN models, and the implications for SN physics and SNR observations. 

\subsection{An explosion with a companion}

One obvious feature of the \OneExp\ and \TwoExp\ SNRs is the shadow cone from the companion, which is a large scale and long-lived feature, visible on the slices and in projection. This is the same effect as we already observed for the D$^6$ model of \cite{Tanikawa2018Three-DimensionalCompanion}, as reported in \cite{Ferrand2022_D6}. This appears to be a robust feature from the presence of a companion. Even though the structure of the inner ejecta may vary, this happens in both the \OneExp\ and \TwoExp\ cases, since, regardless of its final fate, the secondary is always present nearby when the primary explodes. Note that this is not even specific to the double degenerate case, since a similar shadowing is generated by a (farther but larger) non-degenerate companion star (\citealt{Garcia-Senz2012IsRemnants}; Luo et al 2025, in prep). 

Recently \cite{Prust2025EjectaWakes} computed the evolution of the SNR following an ejecta-companion interaction event in both the double-degenerate and single-degenerate cases. The double-degenerate case was obtained from their numerical simulation, while the single-degenerate case was obtained from a fit of previous work by \cite{Kasen2010SeeingCollision}. Similar to our approach, the SNR evolution was calculated in a second step in an expanding grid. The authors focus the analysis on the conical shadow generated by the companion, which they refer to as the ejecta wake. They corroborate our previous findings: the FS stays spherical while the RS is deformed, traveling faster inward along the wake, and RT growth is much enhanced at the edge of the wake. A~difference with \cite{Garcia-Senz2012IsRemnants} and \cite{Ferrand2022_D6} is that they obtain a more asymmetric core at late times, which may be showing us the variations one can get from using different initial binary setups. In any case the simulations by \cite{Prust2025EjectaWakes} confirm that the SNR exhibits observable asymmetries from the presence of a companion for thousands of years. 

\subsection{A~nested explosion}

Turning to the case of a nested explosion, we elucidated that the structure of the ejecta (primary + secondary) is set immediately after the explosion, and so remains preserved for as long as they can be in free expansion. It is a situation that, to our knowledge, has not received much consideration yet. 

One prediction of double degenerate double detonation models, where the secondary WD survives, is the existence of hypervelocity WDs, since the surviving WD keeps its orbital speed of thousands of kilometers per second. Although some hypervelocity WDs have been found \citep{Shen2018ThreeSupernovae}, the currently known number is consistent with only $\sim$2\% of SNe~Ia producing them \citep{Shen2025Hypervelocity}. This means that if double degenerate systems are to make up a large fraction of SNe~Ia, then most of them must explode both WDs. It is thus relevant to look at the resulting SNR. One may have expected a very different morphology for a double explosion, but we have shown that the overall geometry is still quite spherical, since both explosions are rather symmetrical in isolation (except for the shadow effect) and they explode in very close proximity (within a few seconds of ejecta travel time). In fact, the shadow is more evident when the secondary does not explode, so that in a sense the \OneExp\ case is actually more asymmetric when examined globally. 

Detailed high-resolution X-ray imaging will reveal that the \TwoExp\ case has a more complicated ejecta structure. Spatial abundances for \TwoExp\ will be unlike \OneExp, or any other SN model with a single explosion.
One feature that is specific to the \TwoExp\ SNR is the collision of the secondary ejecta with the reverse shock generated by the primary ejecta. Since the collision is asymmetric, it initially produces bright spots on the SNR. Such a feature on its own would be degenerate with the interaction with a non-uniform ISM, although signatures are visible in high-mass elements that are not expected in the shocked ISM. The interaction between the primary and the secondary ejecta is an intrinsic factor of asymmetry for the \TwoExp\ SNR, even if it propagates in a uniform ISM. 

If we could observe the SNR for long enough, over a few hundreds of years, it would become clear how the inner (secondary) ejecta progressively shine \emph{through} the outer (primary) ejecta. Alternatively, if one were able to observe the inner, cold, un-shocked ejecta of a young SNR at other wavelengths, such as optical, infrared, or radio, one could directly see the peculiar layering of the ejecta induced by the nested explosion. Also, the \OneExp\ ejecta are expected to have a global velocity shift, which is not the case for the \TwoExp\ ejecta.

\subsection{Variance in the explosion}

Our work emphasizes again how the initial conditions can be imprinted on the SNR for a very long time. Since we have only one realization of each SN (here in two cases), it is natural to ask how much variance is allowed in the model, especially for \TwoExp. 

A~key aspect is that the secondary explosion is weaker than the primary explosion and happens inside it. It follows that the secondary ejecta are a disturbance to the density profile, but do not change the overall dynamics. In a binary system with unbalanced masses the expectation is that mass transfer happens from the secondary to the primary and thus the primary becomes unstable and explodes -- and then possibly the secondary as well from the impact. If somehow the lower-mass secondary were to explode first, and the higher-mass primary second, the primary explosion would remain the stronger one and so would wipe out the secondary ejecta. This is akin to the case of an explosion in a dense circumstellar medium, e.g. from a stellar wind, a case already considered before (for the case of thermonuclear SNRs, see for instance \citealt{Chiotellis2013Modellingremnant}). The scenario that is possible and for which new simulations may be warranted, is the case when the primary and the secondary are of similar masses. In such a case the interaction between the secondary ejecta and the primary ejecta may be more complex, leading to the interesting question on whether they would merge into a more unified ejecta. 

Another key parameter that determines the relationship between the two explosions is the time delays between the different (four) detonations. However these are set by the geometric configuration of the binary at the moment when the accretion flow can produce an ignition (of the outer layer of the primary). We anticipate it may be difficult to produce markedly different scenarios, like the secondary WD exploding before the primary ejecta have engulfed it, or the secondary WD exploding by the time the primary ejecta are already significantly diluted. Speaking of the accretion, we note that some peculiar localized features of the nested explosion that we pointed out in the SNR, like the enhanced equatorial belt or the inverted RT finger, may be revealing the geometry of the progenitor system, namely the orbital plane and the accretion stream, respectively.

As for the \OneExp\ case, comparison with our similar D$^6$ model may be used to investigate differences in the theoretical modeling of double explosions. The maps are overall in qualitative agreement. The angular scales and their variance are also comparable, although the FS of D$^6$ is more deformed than the one of \OneExp\ (or of \TwoExp).

\subsection{SNRs of interest}

We end this discussion by referring to the list of possible target SNRs established in \cite{Ferrand2022_D6} (section~4.3). The list includes G1.9+0.3, G120.1+1.4 (Tycho’s SNR, SN 1572), G4.5+6.8 (Kepler’s SNR, SN 1604), G327.6+14.6 (SN 1006) in the Galaxy, as well as SNR 0509–67.5, SNR 0519–69.0, and SNR 0509–68.7 (N103B) in the Large Magellanic Cloud. A~recent example of using a young SNR as a probe of the SN physics is the optical detection by \cite{Das2025Calcium} of a double shell of shocked calcium in SNR 0509–67.5, a peculiar feature which is expected from double detonation models. Recent works by Mandal et al. (\citeyear{Mandal2025SignatureTycho}, \citeyear{Mandal2025SNR0509-67.5}) show how the smaller scale structure of SNR images can be used to put constraints on the explosion mechanism, with Tycho and SNR 0509–67.5 found to likely be from a sub-Chandrasekhar mass WD. It remains to be seen whether some unusual, hitherto unexplained observations could be explained as a nested explosion.  

Our mock images provide a first qualitative view of the expected morphological signatures and differences between \OneExp\ and \TwoExp\ SNRs, along different projection axes. 
While this paper is mostly theoretical, observational studies focused on the ejecta dynamics are being conducted by members of our group, in particular on Tycho in X-rays (Godinaud et al, in prep) and on SNR 0509–67.5 in the optical (Das et al, in prep).

\section{Conclusion} 
\label{sec:conclusion}

In this study we have followed a double-degenerate double-detonation SN model into the SNR phase, up to 1500~yr after the explosion. For the first time we consider the case where both the primary and the secondary explode, which, given the time delay between the two explosions of a few seconds, results in a peculiar nested structure. The ejecta of the secondary are expanding inside, and interacting with, the ejecta of the primary, which are themselves interacting with the ISM (that we assumed uniform for simplicity). We have analyzed the structure of the resulting SNR using a variety of representations: 2D slices, 2D projections, 3D contours and their angular variations. We estimated the projected thermal X-ray emission via a proxy.

First, in line with our previous work \citep{Ferrand2022_D6}, we confirm the important role of the companion of the exploding WD, which is here another WD -- whether it also explodes or not. The secondary WD induces a conical shadow in the ejecta of the primary WD, which remains visible well into the SNR phase. The conical shadow intersects the SNR shell in a circular pattern that, depending on the viewing angle, may appear in projection as a ring, an ellipse, or a bar.
This feature should be detectable in the X-ray emission from young SNRs.

A novel aspect of this work is the first comparison between the case when only one WD explodes (\OneExp) and the case when both WDs explode (\TwoExp). This sheds light on what differences may or may not be visible in the SNR phase. Given that the second explosion is significantly weaker, the global dynamics are similar (expansion is somewhat faster for \TwoExp), so the overall morphology is not very different as seen from the outside. Basically, in the \TwoExp\ case, the secondary ejecta stay within the shell generated by the primary ejecta, and they collide into the RS generated by the primary ejecta entering the ISM. At the time of impact this is visible as an over-density, which enhances the thermal emission from the shocked ejecta, locally by a factor of up to about ten. Furthermore, the matter that is revealed during this collision is from the outer layers of the secondary ejecta, which are embedded inside the layers of primary ejecta. After a few hundred years one can see intermediate-mass elements that would otherwise not be present so deep in the ejecta. This peculiar ordering is unique to the \TwoExp\ case. It should thus be possible to detect the nested structure of a double explosion using spatially-resolved spectroscopy in X-rays. This can be pursued using data from existing missions such as \textit{Chandra} and \textit{XMM-Newton}, and \textit{XRISM}, and in the 2030s from next-generation missions such as \textit{AXIS} \citep{Reynolds2023AXIS} and \textit{newATHENA} \citep{Cruise2025NewAthena}. 

This paper is part of a series aimed at elucidating the theoretical signatures of different SN models in the SNR phase. In a forthcoming paper (Fujimaru et al, in prep), using a new, hybrid, faster simulation pipeline we will present a more detailed analysis of the X-ray properties of the SNR, as well as a comparison of the various SN models we have investigated so far.
As for the previous SN models, the extent of variance within the SN model remains to be investigated. In the \TwoExp\ scenario the key parameters are the relative strength of the explosions and the time delay between the explosions. This will be the subject of future work. Another significant complication when comparing models with observations is the role of the local ISM. To improve our modeling of the initial conditions, we will also investigate the role of a 3D circumstellar medium for thermonuclear supernovae.

\begin{acknowledgments}
H.L. acknowledges support from JSPS grant No. JP19K03913.
S.S.H. acknowledges support from the Natural Sciences and Engineering Research Council of Canada (NSERC) through the Canada Research Chairs and Discovery Grants Programs. Supercomputing support in Canada is provided by the Digital Research Alliance of Canada.
S.N. is supported by JSPS Grant-in-Aid Scientific Research (KAKENHI) (A), Grant Number JP25H00675, KAKENHI (B), Grant Number JP23K25874, and JST ASPIRE Program “RIKEN-Berkeley mathematical quantum science initiative”. 
The work of F.K.R. is supported by the Klaus Tschira Foundation, the Deutsche Forschungsgemeinschaft (DFG, German Research Foundation) -- RO 3676/7-1, project number 537700965,
and by the European Union (ERC, ExCEED, project number 101096243). Views and opinions expressed are, however, those of the authors only and do not necessarily reflect those of the European Union or the European Research Council Executive Agency. Neither the European Union nor the granting authority can be held responsible for them. 
A.D. acknowledges support from the Centre National d'Etudes Spatiales (CNES).
D.J.P. acknowledges support from the {\sl Chandra} X-ray Center, which is operated by the Smithsonian Institution under NASA contract NAS8-03060. 
\end{acknowledgments}

\facilities{Simulations were performed on the iTHEMS clusters at RIKEN, and on the Narval supercomputer of the Digital Research Alliance of Canada.}

\software{HEALPix \citep{Gorski2005HEALPixSphere}, SciPy \citep{Virtanen2020SciPyPython}, Matplotlib \citep{Hunter2007MatplotlibEnvironment}}

\appendix

\section{Time evolution of the supernova}
\label{sec:SN_movie}

\begin{figure}[h]
\centering
\includegraphics[width=0.9\textwidth]{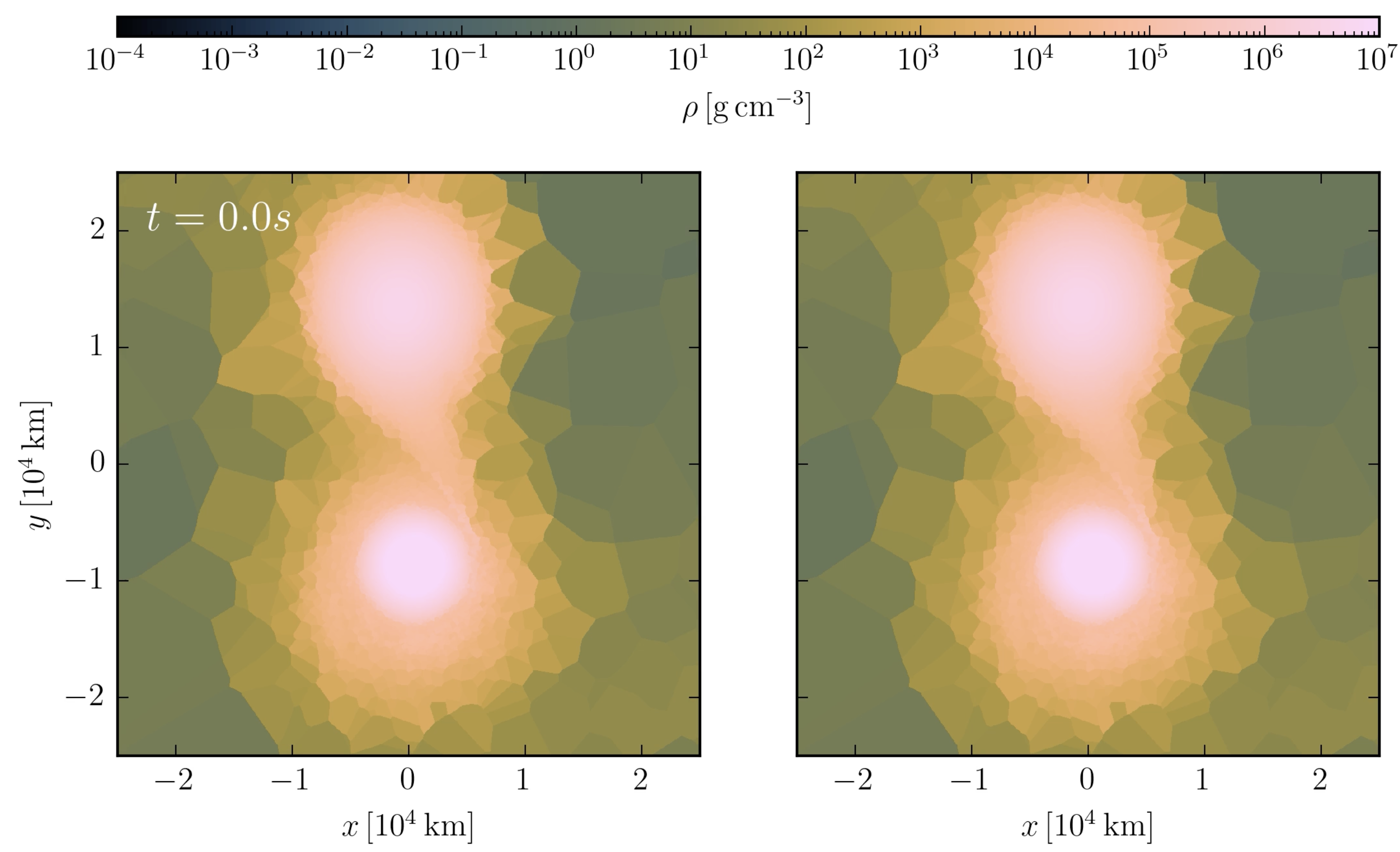}
\caption{Mass density in the SN simulation, in the mid-plane along the $z$ axis. An animated version of this figure is available in the online journal, running up to 150~s for \OneExp\ and 133~s for \TwoExp. Note the varying size of the box as a function of time. The colour scale (ranging logarithmically over 11 orders of magnitude) is shared by the two maps and at all times. 
\label{fig:movie_OneTwoExp}}
\end{figure}

We show in Figure~\ref{fig:movie_OneTwoExp} an animated version of data previously presented in Figure~2 of \cite{Pakmor2022FateSecondary}. In the early frames one can distinguish the accretion stream between the primary and the secondary, as well as the spiraling motion of the WDs. Within a few seconds all this is engulfed by the exploded primary WD. For the \TwoExp\ case, the disruption of the secondary WD is visible starting at 7~s, a double-shock structure from the collision with the primary ejecta develops within a few seconds, and by 20~s the new ejecta no longer show relative motion. The secondary ejecta very quickly settle within the primary ejecta, and by the end of the SN simulation the ejecta form a single structure in near homologous expansion.

From their brutal deceleration in the dense primary ejecta, the secondary ejecta quickly develop the typical RT pattern between the shocks. We note that there exists one inverted RT finger pointing inwards. This suggests that the secondary ejecta encountered a denser obstacle at this location. This happens in a direction close to the edge of the shadow cone but only in one direction so it is not caused by the shadow. We speculate that this corresponds to the endpoint of the spiral tail of matter that is visible in the wake of the secondary before it explodes.

\bibliography{references}{}
\bibliographystyle{aasjournal}

\end{document}